\documentclass[aps, pra, groupedaddress, superscriptaddress, amsmath, amssymb, twocolumn, longbibliography]{revtex4-2}

\usepackage{graphicx}  
\usepackage{xfrac}
\usepackage[normalem]{ulem}
\usepackage{verbatim}   
\usepackage{lipsum}
\usepackage{braket}  
\usepackage{ulem}
\usepackage[dvipsnames]{xcolor}
\usepackage{float}
\usepackage{qcircuit}
\usepackage{circledtext}
\usepackage{rotating} 
\usepackage{orcidlink}
\usepackage{hyperref}
\usepackage[T1]{fontenc}

\begin{document}

\title{ Comparison of spin-qubit architectures for quantum error-correcting codes }

\author{Mauricio Gutiérrez \orcidlink{0000-0002-0826-3521}}
\email{mauricio.gutierrez\_a@ucr.ac.cr}
\affiliation{Escuela de Química, Universidad de Costa Rica, San José 2060, Costa Rica}%

\author{Juan S. Rojas-Arias \orcidlink{0000-0003-3669-0288}}
\affiliation{RIKEN Center for Quantum Computing, RIKEN, Saitama 351-0198, Japan}%

\author{David Obando}
\affiliation{Escuela de Física, Universidad de Costa Rica, San José 2060, Costa Rica}%

\author{Chien-Yuan Chang \orcidlink{0000-0003-0548-7391}}
\email{cychang@ee.nthu.edu.tw}
\affiliation{Department of Physics and Center for Quantum Science and Technology, National Tsing Hua University, Hsinchu 30013, Taiwan}
\affiliation{National Center for Excellence in Quantum Information Science and Engineering, National Tsing Hua University, Hsinchu 30013, Taiwan}

\begin{abstract}

We investigate the performance of two quantum error-correcting codes--the surface code and the Bacon-Shor code--for implementation with spin qubits in silicon. In each case, we construct a logical qubit using a planar array of quantum dots, exploring two encoding schemes: one based solely on single-electron Zeeman qubits (Loss-DiVincenzo qubits) and a hybrid approach combining Zeeman and singlet-triplet qubits. For both codes, we evaluate key performance metrics, including logical state preparation fidelity and cycle-level error correction performance, using state-of-the-art experimental parameters. Our results show that the hybrid encoding consistently outperforms the pure Zeeman qubit implementation. By identifying the dominant error mechanisms that limit quantum error correction performance, our study highlights concrete targets for improving spin-qubit hardware and provides a path toward scalable fault-tolerant architectures.  In particular, we find that the logical error rate is not limited by memory errors but rather by gate errors, especially $1$- and $2$-qubit gate errors. 

\end{abstract}

\maketitle

\section{Introduction}

A large-scale quantum computer will provide revolutionary information processing capabilities in various fields \cite{shor1999polynomial, farhi2001quantum, alan2005, biamonte2017quantum, Lloyd96}.  The construction of such a device will require a hardware platform with sufficiently long qubit coherence times, precise and fast qubit control, and the ability to scale up easily \cite{DiVincenzoCriteria}.  Spin qubits in semiconductor quantum dots (QD) \cite{hanson2007spins, zwanenburg2013silicon, burkard2023semiconductor} constitute a very promising alternative to other solid-state systems.  They offer great scalability, compatibility with existing highly advanced semiconductor fabrication technologies \cite{zwerver2022qubits, li2020flexible}, and the possibility of parallel qubit control \cite{Veldhorst2017, john2024bichromatic, lawrie2023simultaneous}.  Furthermore, spin qubits in silicon-based systems have also achieved low error rates for single-qubit and entangling gates, as well as measurements \cite{noiri2022fast, xue2022quantum, mills2022two, wu2024hamiltonian, tanttu2024assessment}. 

Running deep quantum algorithms that provide quantum advantage will require qubit error rates below $10^{-10}$ \cite{campbell2021early, kivlichan2020improved}.  These demandingly low values will most likely never be achieved on physical qubits and instead will require logical qubits protected by quantum error correction (QEC) \cite{ShorCode, TheoryQECC, Shor96_1, Calderbank97, KitaevAnyons, TerhalReviewQEC}. A QEC code encodes logical qubits non-locally across multiple physical qubits to protect against imperfections in control pulses and environmental noise. Crucially, this protection is effective as long as the error rates remain below certain thresholds. According to the threshold theorem, if the physical error rates lie below these thresholds--which depend on the choice of QEC code, the details of the correction circuitry, and the characteristics of both the noise and qubit connectivity--then logical error rates can be exponentially suppressed by increasing the code size, with only polylogarithmic overhead in resources \cite{Aharonov96, KnillThreshold, PreskillThreshold, Burkard2005, Nielsen2005, Leung2006, AliferisThreshold}.  Demonstrating a logical encoded qubit is a crucial milestone on the exciting quest towards fault-tolerant (FT), large-scale quantum computing \cite{acharya2024quantum}.

Although still behind QEC experimental demonstrations by other platforms like superconductors \cite{acharya2024quantum}, trapped ions \cite{Quantinuum2024}, and neutral atoms \cite{Bluvstein2024}, several proof-of-concept QEC experiments have been showed recently on spin qubits \cite{takeda_quantum_2022, van2022phase}. These experiments have focused on repetition codes capable of correcting phase flips.  A logical next step would be to implement a distance-$3$ fully quantum code resilient not only against dephasing, but also capable of correcting any single-qubit error. It will also be crucial to achieve a break-even point where a logical, error-corrected qubit exhibits a lower error rate than its constituent physical qubits.   

To guide future QEC experimental demonstrations on spin qubits, we study the implementation of two distance-$3$ QEC codes on a silicon spin qubit platform.  We focus on two QEC codes compatible with 2D architectures and nearest-neighbor connectivity, namely the surface code \cite{KitaevSurface, Dennis2002} and the Bacon-Shor (BS) code \cite{BSOriginal}.  Although the former is arguably the most famous QEC code, the latter has received much less attention and, to the best of our knowledge, has never been implemented on a solid-state system.  

We also compare two different qubit encoding schemes.  The electron spin is a native two-level system for qubit encoding \cite{loss1998quantum}, known as the Loss-DiVincenzo (LD) qubit. Alternatively, it is possible to use two electrons in separate QDs to encode the qubit in their joint spin state, known as the singlet-triplet (ST) qubit \cite{petta2005coherent}.  We compare a layout where all qubits are implemented with the LD encoding with a hybrid one where the LD qubits are used for the data and ST qubits are used for the ancilla, motivated by the longer coherence time of the LD qubits but the considerably faster readout of the ST qubits \cite{noiri2020radio, takeda2024rapid}. Combining different encodings may help to exploit their strengths and mitigate their limitations \cite{mehl2015simple, noiri2018fast}.    

For the two QEC codes (surface and BS) and the two alternative qubit encoding schemes (all-LD and hybrid), we simulate both a FT QEC step and a FT preparation of logical $|0\rangle$ and $|+\rangle$ states.  We perform stabilizer simulations with a circuit-based error model that includes gate, measurement, and memory errors, with state-of-the-art values.  We find that the hybrid encoding scheme considerably outperforms the all-LD scheme, mainly because of the ST shorter readout duration, which reduces decoherence on the idle qubits.  The surface code results in a slightly lower logical failure rate than the BS code during the FT QEC step.  However, the BS code offers an improvement of almost $2$ orders of magnitude for the preparation of logical $|0\rangle$ and $|+\rangle$ states, provided that the layout connectivity allows for direct entangling gates between neighboring data qubits along the same row, which open the possibility to coherent (as opposed to projective) logical state preparation.

We use an error-subset sampler to explore which noise parameters are more critical to reduce the logical failure rate.  The logical error rate is dominated by gate errors and not memory errors.  In particular, we find that a decrease in the infidelity of $1$- and $2$-qubit gates would have the largest impact on the logical error rate.    For the ST qubit, a longer readout integration time has the conflicting effect of both decreasing the readout infidelity and increasing the qubit decoherence due to waiting.  We numerically find readout integration times that result in the lowest logical failure rates.  These optimal integration times are not equal to the integration time that results in the lowest readout infidelity.  Similar results have been also found recently by Het\'enyi and Wootton \cite{hetenyi2023tailoring}.  For the state-of-the-art integration time, the hybrid encoding scheme can outperform a bare physical qubit (without spin-echo protection) for both QEC codes.  However, the logical error rates hit a floor on the order of $10^{-2}$, which is limited by gate infidelities and not memory errors.     

The article is organized as follows. Section \ref{sec:PhysImp} provides a description of the encoding schemes and the physical system layout.  Section \ref{sec:QEC} gives an overview of the two QEC codes. Section \ref{sec:SimDetails} explains the details of the simulation, including the noise model, the circuits that were simulated, and the error-subset sampler.  Section \ref{sec:Results} presents and analyzes the results before concluding in Section \ref{sec:Conclusions}.

\section{Physical implementation} \label{sec:PhysImp}

\begin{figure}
    \vspace{-7pt}
    \centering
    \includegraphics[width=0.95 \columnwidth]{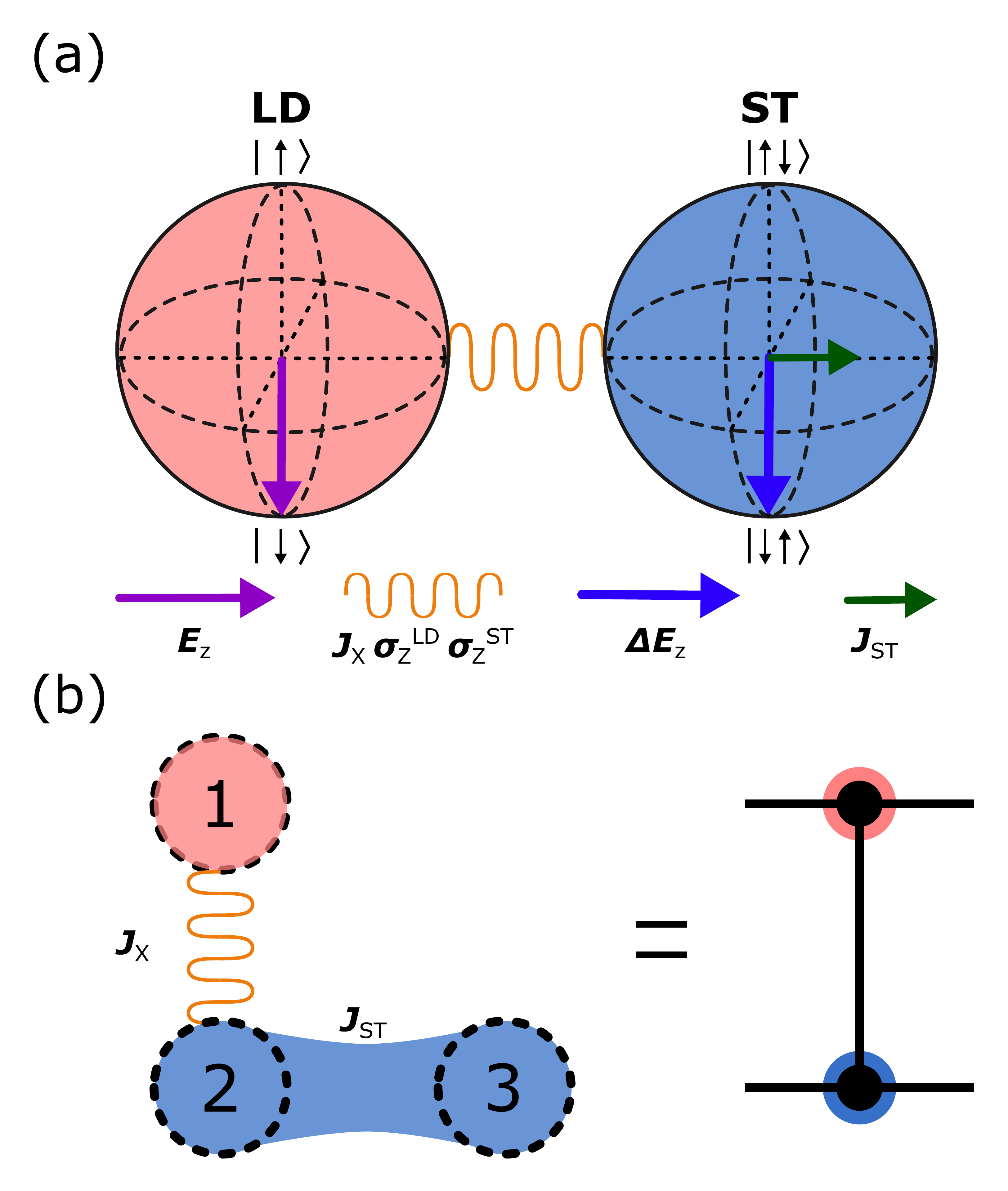}       
    \caption{(a) Illustration of \textbf{LD} (pink) and \textbf{ST} (blue) qubits Bloch spheres, highlighting distinct color-coded interactions. (b) This study employs two spin qubit encodings: an \textbf{LD} qubit (pink) realized by a single electron spin in QD1, and an \textbf{ST} qubit (blue) encoded in the two-spin wavefunction across QD2 and QD3. The inter-qubit exchange coupling ($J_X$) mediates the interaction between QD1 and QD2, while the intra-qubit exchange coupling ($J_{ST}$) governs the interaction within the ST qubit (QD2 and QD3). A controlled-phase ($Z$) gate can be implemented by exploiting the exchange coupling ($J_X$) between encodings \cite{mehl2015simple, noiri2018fast, nakajima2017robust}. }
    \label{state}
\end{figure}

The platform of interest consists of qubits encoded in the spin states of electrons confined in semiconductor QDs \cite{burkard2023semiconductor}. Typical devices are based on semiconductor heterostructures featuring a quantum well or a semiconductor–oxide interface that supports the formation of a two-dimensional electron gas (2DEG)\cite{hanson2007spins}. Lateral confinement of electrons into QDs is achieved by applying voltages to lithographically defined metallic gates patterned on the surface of the chip.

Within this platform, we investigate two types of spin qubit encodings. The first is the LD qubit, where the computational basis is defined by the spin projection along the magnetic field axis of a single electron confined in a QD \cite{loss1998quantum}. Specifically, the qubit states are $\ket{0_\mathrm{LD}} \equiv \ket{\uparrow}$ and $\ket{1_\mathrm{LD}} \equiv \ket{\downarrow}$, forming the Bloch sphere shown in pink in Fig.~\ref{state}a. The energy splitting between these states is the Zeeman energy $E_Z = g\mu_B B_z$, where $g$ is the material-dependent $g$-factor, $\mu_B$ is the Bohr magneton, and $\vec{B} = B_z \hat{z}$ is an applied static magnetic field.

Control of the LD qubit is achieved via spin rotations on the Bloch sphere, implemented either by electron spin resonance (ESR) using an oscillating magnetic field perpendicular to $\vec{B}$ \cite{koppens2006driven}, or via electric dipole spin resonance (EDSR) using an oscillating electric field in the presence of a magnetic field gradient \cite{pioro2008electrically}. In both cases, the drive frequency must match the qubit transition frequency $E_Z/\hbar$. Readout is typically performed via energy-selective tunneling: the Fermi level of a nearby reservoir is tuned so that only the $\ket{\uparrow}$ state can tunnel out of the dot. The spin state is thus mapped to a charge signal, detected through the presence or absence of an electron tunneling event \cite{elzerman2004single}.

The second encoding of interest is the singlet–triplet (ST) qubit, defined by the spin states of two interacting electrons in a double quantum dot \cite{levy2002universal,petta2005coherent}. The electrons interact through the exchange coupling $J_{ST}$, and in the presence of a magnetic field gradient, the Zeeman {energy} at each dot differs by $\Delta E_Z$. The relevant Hilbert space consists of four two-electron spin states. The computational subspace is formed by the two odd-parity eigenstates: a singlet-like state $\ket{0_\mathrm{ST}} \equiv a_+ \ket{\uparrow\downarrow} - a_- \ket{\downarrow\uparrow}$ and a triplet-like state $\ket{1_\mathrm{ST}} \equiv a_- \ket{\uparrow\downarrow} + a_+ \ket{\downarrow\uparrow}$, where the coefficients are given by $a_\pm^2 = \tfrac{1}{2} \left(1 \pm \Delta E_Z / \sqrt{\Delta E_Z^2 + J_{ST}^2} \right)$. These states form the Bloch sphere depicted in blue in Fig.~\ref{state}a. The other two eigenstates, $\ket{\uparrow\uparrow}$ and $\ket{\downarrow\downarrow}$, are even-parity leakage states that lie outside the computational subspace. Here, $\ket{m_1 m_2}$ with $m_i=\{\uparrow,\downarrow\}$ denotes the spin state of electrons in the left and right dots, respectively.

In the absence of a magnetic gradient, the computational basis corresponds to the pure singlet and $T_0$ triplet states with $a_\pm = 1/\sqrt{2}$. A nonzero $\Delta E_Z$ hybridizes the singlet and triplet states. In the limit $\Delta E_Z \gg J_{ST}$, the basis states become product states: $\ket{\Tilde{0}_\mathrm{ST}} \simeq \ket{\uparrow\downarrow}$ and $\ket{\Tilde{1}_\mathrm{ST}} \simeq \ket{\downarrow\uparrow}$. Gate voltages allow for the tuning of both $J_{ST}$ and, to some extent, $\Delta E_Z$, meaning the qubit eigenstates can vary depending on device-specific tuning. The energy splitting between qubit states is $E_\mathrm{ST} = \sqrt{\Delta E_Z^2 + J_{ST}^2}$.

Qubit control is achieved via either resonant modulation of the exchange interaction \cite{takeda2020resonantly} or by adiabatic rotations that exploit the differing effects of $J_{ST}$ and $\Delta E_Z$ on the qubit Hamiltonian—they act along orthogonal axes in the Bloch sphere \cite{Wu2014two,berritta2024real}, shown by blue and green arrows in Fig.~\ref{state}a.
Readout is typically performed via Pauli spin blockade (PSB) \cite{petta2005coherent}. The system is pulsed to a configuration where interdot tunneling is energetically allowed only for the singlet state. In this readout configuration, transitions from the (1,1) to (0,2) charge state are permitted for odd-parity (antisymmetric spin) states like the singlet, but are blocked for even-parity (symmetric spin) triplets due to the Pauli exclusion principle. The odd-parity qubit state $\ket{1_\mathrm{ST}}$, which resembles $\ket{T_0}$ in the absence of a field gradient, is typically prevented from tunneling because the (0,2) triplet state is energetically inaccessible. While magnetic field gradients can mix the singlet and triplet states at the time of readout \cite{seedhouse2021pauli}, this can be mitigated through fast spin-to-charge conversion, a strength of PSB-based readout.

Two-qubit gates are implemented by activating the exchange interaction between adjacent quantum dots. For LD qubits, pulsing the tunnel barrier between singly occupied QDs enables a controllable exchange coupling that allows the implementation of entangling gates such as $\sqrt{\text{SWAP}}$ or controlled-phase operations \cite{mills2022two}. In the case of ST qubits, inter-qubit exchange between electrons in neighboring double QDs can produce entangling dynamics when the coupling is tuned to preserve the logical subspace. Hybrid LD–ST qubit architectures also support two-qubit gates via exchange coupling, $J_\mathrm{X}$, across the LD spin and one of the spins in the ST pair \cite{noiri2018fast}. This situation is schematically depicted in Fig.~\ref{state}b. The interaction $J_\mathrm{X}$ results in a conditional evolution of the LD spin that depends on the state of the ST qubit, allowing universal control across mixed qubit types. This unified approach to entangling gates supports fast and local two-qubit operations while maintaining compatibility with heterogeneous qubit layouts.

\section{Quantum error-correcting codes}  \label{sec:QEC}

A QEC code encodes one or more logical qubits as non-local entangled states of a larger number of physical qubits. QEC codes will be a crucial component of a large-scale FT quantum computer since they offer the clearest path to building logical qubits with arbitrarily low error rates.   A $[[n,k,d]]$ QEC code refers to a code composed of $n$ physical qubits that encodes $k$ logical qubits with a distance $d$.  The distance refers to the minimal number of discrete errors that map one logical state to its orthogonal logical state.  A distance-$d$ code can detect $d-1$ errors, but can only correct $\lfloor \frac{d-1}{2} \rfloor$ errors \cite{Dennis2002}.

Stabilizer codes constitute one of the most common QEC code families.  For $n$ physical qubits, these codes are defined by an Abelian group of Pauli operators spanning the $2^n$-dimensional Hilbert space, known as the stabilizer group. The logical codespace of a stabilizer code is the simultaneous (+1)-eigenspace of all operators within this group. A set of independent stabilizers (non-unique) capable of generating the entire stabilizer group is known as the stabilizer generator. To perform error detection and correction, the eigenvalues of the stabilizer generators (the syndrome) are measured indirectly with ancillary qubits. Afterwards, a decoder is used to infer the error that occurred based on the syndrome \cite{Battistel2023}.  

\begin{figure}[t]
    \centering
    \includegraphics[width=0.95 \columnwidth]{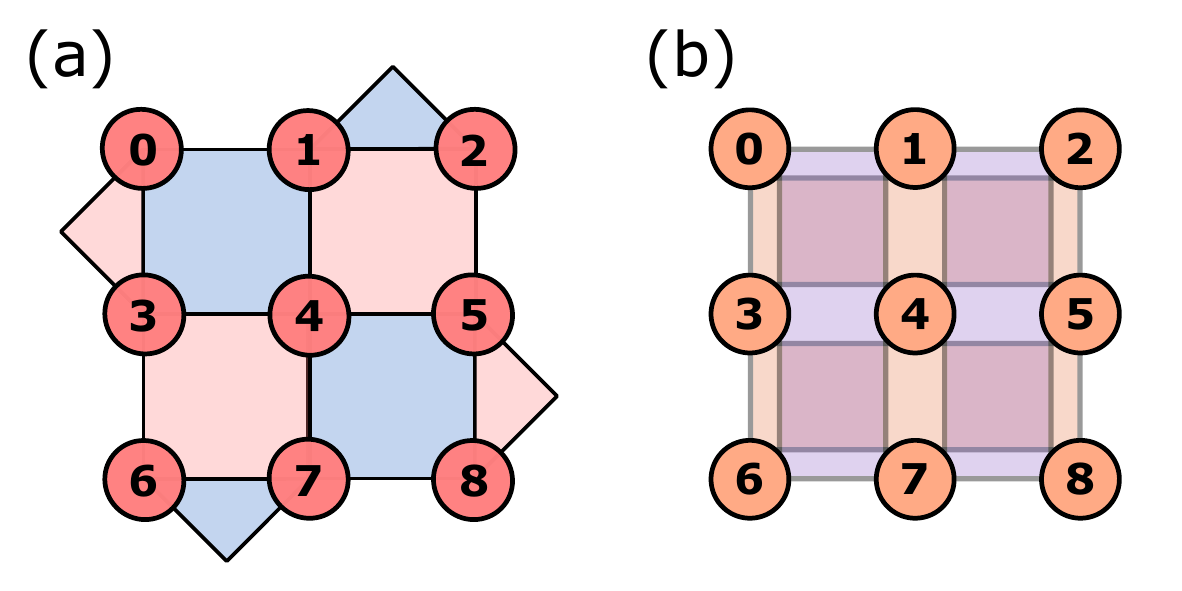} 
 %
\centering
\begin{minipage}{0.27\textwidth}
\vtop{
\centering
\begin{tabular}{|c|c|}
\multicolumn{2}{c}{Surface Code} \\
\hline
$S^{(1)}_X$ & $X_1X_2$ \\
\hline
$S^{(2)}_X$ & $X_0X_1X_3X_4$ \\
\hline
$S^{(3)}_X$ & $X_4X_5X_7X_8$ \\
\hline
$S^{(4)}_X$ & $X_6X_7$ \\
\hline
\hline
$S^{(1)}_Z$ & $Z_0Z_3$ \\
\hline
$S^{(2)}_Z$ & $Z_1Z_2Z_4Z_5$ \\
\hline
$S^{(3)}_Z$ & $Z_3Z_4Z_6Z_7$ \\
\hline
$S^{(4)}_Z$ & $Z_5Z_8$ \\
\hline
\end{tabular}
}
\end{minipage}
\hfill
\begin{minipage}{0.2\textwidth}
\vtop{
\centering
\begin{tabular}{|c|c|}
\multicolumn{2}{c}{Bacon-Shor Code} \\
\hline
$S^{(1)}_X$ & $X_0X_1X_3X_4X_6X_7$ \\
\hline
$S^{(2)}_X$ & $X_1X_2X_4X_5X_7X_8$ \\
\hline
\hline
$S^{(1)}_Z$ & $Z_0Z_3Z_1Z_4Z_2Z_5$ \\
\hline
$S^{(2)}_Z$ & $Z_3Z_6Z_4Z_7Z_5Z_8$ \\
\hline
\end{tabular}
}
\end{minipage}
\label{tab:stacked_stabilizers}

  \caption{Qubit layout for (a) the distance-3 rotated surface code and (b) the distance-3 BS code. In both figures, only the data qubits are labeled. The respective stabilizer generators are shown below.} 
  \label{fig:QECC_layout}
\end{figure}

Figure \ref{fig:QECC_layout} depicts the two distance-3 QEC codes we explore in this paper.  They are both Calderbank-Shor-Steane (CSS) codes.  This implies that they have a stabilizer generator set composed of operators which are exclusively composed of $X$ or $Z$ Pauli matrices.  Notice that this stabilizer generator set is not unique, but it is one that guarantees locality (each operator acts on neighboring qubits).

\subsection{Surface code}

Two-dimensional (2D) rotated surface codes \cite{KitaevSurface, Dennis2002, fowler2012surface} are a family of $[[d^2,1,d]]$ stabilizer codes defined on an array of $n=d^2$ physical data qubits on a plane.  The stabilizer group is generated by a set of $(n-1)$ independent Pauli operators that act locally on the physical qubits.  This allows for the implementation of the code with only nearest-neighbor interactions and makes it amenable for solid-state platforms. These generators are of weight 4 if they act on the bulk of the lattice or of weight 2 if they act on the edges, where the weight refers to the number of data qubits involved.  The additional ancillary physical qubits are used to measure the $(n-1)$ generators in a quantum non-demolition (QND) fashion in order to project the logical qubit to the stabilizer eigenspace, as well as detect and correct errors.  Planar surface codes are currently one of the leading candidates to achieve large-scale fault tolerance due to their high error threshold \cite{fowler2009highthreshold} and their relatively mild requirement of only nearest-neighbor interactions on a 2D plane. For instance, a surface-17 logical qubit with the hybrid encoding could be implemented in a connectivity-4 architecture as shown in Fig.~\ref{fig:layout_two_codes}b. While implementing 2D surface code in a spin-based system requires advanced manufacturing, such a layout could be realized in silicon \cite{unseld2024baseband} and other semiconductor materials \cite{mortemousque2021enhanced, rao2025modular, borsoi2024shared}.

Apart from their high error threshold and local connectivity, planar surface codes offer the additional advantage of allowing the transversal implementation of several logical gates, including all Pauli operators, CNOT, and, up to a permutation of the qubits' labels, Hadamard.  Transversality refers to the property of a logical operation of acting on a separate set of physical qubits for each logical qubit.  For example, a logical $X_L$ gate is transversal for the surface code because it can be implemented as a product state of physical $X$ gates on each physical qubit ($X ^{\otimes n}$).  Transversality guarantees error containment because errors that occur on one physical qubit will not propagate to the other physical qubits in the same logical qubit.  Transversal logical gates are highly desirable since they are naturally fault-tolerant.  However, for the surface code, the logical gate $T$ ($R_Z(\pi/4)$) is not transversal and requires a special protocol to be implemented in a fault-tolerant fashion, most commonly magic state distillation \cite{MSDOriginal, Litinski2019magicstate}.  There have been several fault-tolerant experimental implementations of the rotated surface code, including distance-3 \cite{Krinner2022, zhao2022realization}, distance-5 \cite{google2023suppressing}, and distance-7 \cite{acharya2024quantum}.  All of these have been on superconducting systems.

\subsection{Bacon-Shor code}

Like 2D surface codes, 2D BS codes \cite{BSOriginal, ShorCode, BSAliferisCross} are a family of $[[d^2,1,d]]$ CSS codes defined on a planar array of $n=d^2$ physical data qubits.  In contrast to surface codes, BS codes are subsystem codes \cite{OQEC1, OQEC2}.  This implies that the logical subspace has dimensions higher than 2 and thus contains several logical qubits, which can be seen as subsystems of the codespace.  Of these subsystems, typically only one is chosen as the working logical qubit.

For symmetric or square 2D BS codes, there are $(d-1)$ X stabilizer generators and $(d-1)$ Z stabilizer generators, each of weight $2d$.  $X$($Z$) stabilizer generators correspond to vertical (horizontal) rectangles as shown in Fig.~\ref{fig:QECC_layout}b for the distance-$3$ case.  The reduced number of stabilizers implies that there is more than one logical qubit.  However, only one of these has the full code distance $d$ and is used as the actual logical qubit.  The rest are gauge (logical) qubits that are not used to store information.  For example, for the distance-$3$ planar BS code shown in Fig.~\ref{fig:QECC_layout}b, there are $5$ encoded qubits, of which one is the working logical qubit and the other $4$ are the gauge qubits.

If the gauge logical qubits are fixed to definite states, we obtain a subspace code.  Different gauge fixings result in different subspace codes.  For example, if we fix all the gauge qubits to the state $|0 \rangle$, we obtain the original Shor code \cite{ShorCode}.  Alternatively, if we fix two gauge qubits to $|0 \rangle$ and the other two to $|+ \rangle$, we obtain the rotated distance-$3$ surface code described previously.  All of these codes are instances of the more general compass code family \cite{CompassCodes}.  In contrast to the surface code, the BS code does not present a quantum error correction threshold as the lattice size increases \cite{BSNoThreshold}, although a threshold can be guaranteed by different techniques \cite{BSAliferisCross, LessBacon, BSBoardGames}.  



The BS code has several very useful properties. The high-weight stabilizers can be measured indirectly by breaking them into products of weight-$2$ gauge operators (or some combination of them) and calculating their total parity \cite{BSAliferisCross}.  This scheme is very convenient for architectures with only nearest-neighbor connectivity, such as solid-state systems, and it is the approach proposed in this paper. Also, just like for the surface code, all logical Pauli operators, logical CNOT, and logical Hadamard can be implemented transversally on the BS code.   Finally, for the distance-$3$ BS code analyzed in this paper, it is possible to prepare logical $|0\rangle_L$ and $|+\rangle_L$ states in a FT way by only creating $3$ Greenberger-Horne-Zeilinger (GHZ) states \cite{Egan2021}.  This offers a promising advantage over the distance-$3$ surface code, which requires $1-2$ rounds of QND stabilizer measurements.   The distance-$2$ and distance-$3$ BS codes have been implemented in trapped-ion systems \cite{422,Egan2021,DukeSteaneEC}, but, to our knowledge, no BS code has been implemented in a solid-state system.



\begin{figure}
    \centering
    \includegraphics[width=1 \columnwidth]{Fig3.pdf} 
       \vspace{0pt}
    \caption{Layouts of QEC codes implemented with the hybrid scheme. The $XXXX$ ($ZZZZ$) parity checks are labeled in blue (pink) following ref. \cite{tomita2014low}. (a) Surface-17 code with data qubits in red and ancilla qubits in blue. (b) QD implementation of the surface-17 code, with data qubits (red single dots) encoded as LD qubits and ancilla qubits (blue pairs) encoded as ST qubits. The connectivity of each single quantum dot is 4.  (c) BS-17 code with data qubits in orange and ancilla qubits in purple. (d) QD implementation of the BS-17 code, with data qubits (orange single dots) encoded as LD qubits and ancilla qubits (purple pairs) encoded as ST qubits. The BS-17 implementation requires a qubit connectivity of 6 to allow for the unitary preparation of GHZ states along the rows of the code.}   
   \label{fig:layout_two_codes}
\end{figure}

Figure \ref{fig:layout_two_codes} depicts the layouts of the two QEC codes analyzed under both abstract (left) and a QD settings (right).  For the surface-17 code, the square layout has a qubit connectivity of 4.  However, for the BS-17 code, our proposed layout is hexagonal and requires a qubit connectivity of 6.  This is crucial to directly perform entangling gates between neighboring data qubits on the same row to allow for the coherent (as opposed to projective) logical state preparation in the BS-17 code.  Triangular quantum dot arrays have been experimentally implemented \cite{TriangularMahalu,TriangularTarucha,TriangularPetta}.  These triangular arrays would be the basic unit of a larger hexagonal lattice.  

{ An important assumption in our simulations concerns the readout architecture. The relatively low readout infidelity for the ST qubit reported in Table \ref{tab:parameters} corresponds to charge-sensor-based readout, since dispersive readout currently exhibits significantly higher noise \cite{kam2024fast}. Charge sensing requires close proximity to a sensor, which in a dense 2D array implies that only boundary qubits can be measured directly. In our proposed layouts there is no available footprint to integrate charge sensors within the bulk of the array without compromising qubit density and connectivity. As a result, measuring a bulk qubit would require shuttling the corresponding electron to a boundary QD located next to a charge sensor, where readout and initialization can be performed. For the distance-3 codes analyzed here, however, assuming direct measurement of all ST qubits, including those in the bulk, is not unreasonable. In the surface-17 layout, all ancillary ST qubits are already positioned at the boundary and close to a charge sensor. In the BS-17 layout, only two ancillary ST qubits reside in the bulk, and shuttling these electrons to the boundary involves short distances. In Appendix~\ref{app:noreadoutbulk} we provide initialization and readout schedules that minimize electron shuttling and result in a negligible impact on the logical error rates.  This leaves our results essentially unchanged compared to the idealized case where every qubit can be measured in place.  For larger-distance QEC codes, where most ancilla qubits would lie in the bulk, a detailed scheduling of electron shuttling operations, including their cumulative error contribution, must be included in any realistic scalability analysis.}



\section{Simulation details}  \label{sec:SimDetails}

Our simulations are based on the stabilizer formalism \cite{Gottesmanthesis}, which has the advantage of allowing for an efficient (polynomially costly) implementation on a classical computer \cite{MikeIke}, in contrast to a state-vector or density-matrix simulation, which scales exponentially.  Our simulator is a set of wrapper tools with CHP \cite{CHP} as its core engine and has been employed in previous investigations of quantum error-correcting codes \cite{FaultPaths,Crosscode,IonTrapColin,LogicalCNOT,SteaneECBS}.  It is worth pointing out that CHP is not the only stabilizer simulator and that there are more recent alternatives with an even better performance in terms of speed of simulation \cite{FastSimBriegel, Stim}.  We employ our wrapper tools mainly because of its importance (error subset) sampling capabilities, which are discussed in Subsection \ref{subsec:sampler}. 

Not all gates and noise processes can be simulated in the stabilizer formalism.  For gates, we are limited to preparations of a Pauli basis state (eigenvectors of the Pauli matrices), measurement in a Pauli basis, as well as all gates in the Clifford group, which is generated by Hadamard, CNOT, and S ($R_Z(\pi/2) = \exp(-i \pi Z /4)$).  The Clifford group is not universal for quantum computing in the sense that it does not contain all possible quantum gates.  However, it is sufficient to simulate important subroutines in fault-tolerant quantum computing, in particular quantum error correction of stabilizer codes, which only contain Clifford gates and Pauli preparations and measurements.

\subsection{Noise model}

Noise processes outside the Clifford group need to be approximated with errors in the Clifford group or a subgroup therein, such as the Pauli group.  This can be achieved with procedures like the Pauli twirling approximation \cite{PTADur, PTAGeller, PTABenjamin}.  Furthermore, it is also possible to perform a constrained minimization to obtain an honest Pauli or Clifford approximate channel to the non-Clifford noise process, in the sense that the approximation does not underestimate the deleterious effect of the noise \cite{Magesan, CliffApprox, Puzzuoli, Guti1, Guti2}.      

The noise model employed in our simulations consists of the following stochastic Pauli error channels:

\begin{enumerate}
    \item Bit-flip (X) error after $|0\rangle$ state preparation with probability $p_p$, modeling imperfect initialization.
    \item Bit-flip (X) error before measurement in the $Z$ basis with probability $p_m$, modeling misassignment of the measurement outcome caused by finite signal-to-noise ratio, relaxation during readout, or charge-sensor noise \cite{takeda2024rapid}.
    \item Y error after rotations about the Y axis with probability $p_{1q}$, modeling stochastic over- and under-rotations caused by qubit-frequency fluctuations that detune the qubit from the microwave drive during the pulse. The only single-qubit gates in our circuits are $\pm \pi/2$ rotations about the Y axis.
    
    \item ZZ error after the CZ gate with probability $p_{\textrm{CZ}}$, modeling conditional phase errors arising from fluctuations in the exchange interaction during the entangling operation.
    \item Independent Z errors on every idle qubit during operations on other qubits, accounting for dephasing, with probability $p_\mathrm{idle} = \frac{1}{2} \left( 1- \exp(-(t_\mathrm{idle}/T_{2}^*)^2) \right)$, where $t_\mathrm{idle}$ is the duration of the corresponding operation and $T_2^*$ is the dephasing time constant of the idling qubit. The Gaussian decay characterizes the coherence loss in systems exposed to $1/f$-type (low-frequency) noise, which is the most common in solid-state systems.   
\end{enumerate}

Notice that in a hybrid scheme $p_\mathrm{idle}$ will not be the same for every qubit, since in general the LD qubit and the ST qubit will have different values of $T_2^*$. Likewise, the error probabilities of state preparations, measurements, single-qubit gates, and CZ gate will be different depending on the type of qubit. 

Throughout our simulations, single-qubit gates are assumed to have a duration of exactly 0, so the only operations that have a waiting time and result in idling qubits are CZ gates, measurements, and preparations.   This is not a crass assumption, since, as shown in Section \ref{sec:Results}, most of the dephasing occurs during the ancilla measurements, which constitute the rate-limiting step.  Also, the dephasing effect on the idle qubits during single-qubit gates is equivalent to having longer CZ gates, so it is always possible to increase the duration of the CZ gates to account for the dephasing during single-qubit gates.

We assume that idle-qubit dephasing is uncorrelated between different qubits. While low-frequency noise in solid-state spin qubits can exhibit spatial correlations \cite{yoneda2023noise,rojas2023spatial,rojas2026origins,rojas2026inferring}, such effects are expected to produce at most a modest quantitative change for the short distance-3 protocols considered here. Correlated noise is known to become increasingly important with growing code distance and system size \cite{clemens2004quantum}, and its impact on spin-qubit QEC is beyond the scope of the present work and will be addressed elsewhere.

We note that additional noise mechanisms relevant for spin qubits, such as energy relaxation and valley-related effects, are not modeled explicitly. For both LD and ST qubits in silicon, energy relaxation is subdominant on the timescales considered here, as $T_1$ times are typically much longer than the corresponding dephasing times. For example, for ST qubits $T_1 \simeq 3.3\,\mathrm{ms}$ \cite{takeda2024rapid}, whereas the $T_2^*$ values used in this work are on the order of tens of microseconds (Table~\ref{tab:parameters}). Valley-related imperfections are implicitly incorporated into the effective preparation, measurement, and gate error probabilities listed in Table~\ref{tab:parameters}, which are extracted from experimental reports.

\subsection{Importance sampler}
\label{subsec:sampler}

In stabilizer simulations of noisy quantum circuits, errors are modeled as stochastic occurrences of operations simulable within the stabilizer formalism, typically Pauli operators.  Errors are randomly sampled from the corresponding probability distribution and the resulting noisy circuit is executed to determine if the selected error configuration results in a logical failure or not.  

In the traditional sampler, the whole error set is directly sampled, \textit{i.e.}, the quantum circuit is exhaustively traversed and, after each noisy gate or step, an error is randomly added according to its error probability.  Although widely used, this direct sampler can become very slow for low physical error rates, since it takes many runs to generate an error configuration that will result in a logical error.  To accelerate the simulations, we use an importance (error subset) sampler.  In this approach, the total error set is divided into non-overlapping subsets according to the error weight (number of errors).

For a noise model with $m$ independent error parameters, we label each subset with the vector $\vec{w} = (w_1, w_2, ..., w_m)$, where $w_i$ denotes the number of errors associated with the parameter $i$.  For example, in the simulations performed in the present work, $m=8$.  These correspond to errors (1) after state preparations, (2) before measurements, (3) after single-qubit gates, (4) after two-qubit gates, (5) on idling data qubits during ancilla preparations, (6) on idling data qubits during ancilla measurements, (7) on idling data qubits during 2-qubit gates, and (8) on idling ancilla qubits during 2-qubit gates.  

The subset sampler involves two steps:
\begin{enumerate}
    \item The total probability of occurrence $A_{\vec{w}}$ is computed for the most likely error subsets.  This can be calculated analytically with the following formula:
    \begin{equation*}
        A_{\vec{w}} = \prod_{i=1}^{m} \binom{n_i}{w_i} p_{i}^{w_i} (1-p_i)^{n_i - w_i},
    \end{equation*}
    where $n_i$ ($p_i$) is the total number of gates or steps (the physical probability) associated with the noise parameter $i$.
    \item For the selection of the most likely error subsets, direct sampling is performed to estimate the logical error rate, $p_L^{(\vec{w})}$. 
\end{enumerate}

We then calculate lower and upper bounds to the exact logical error rate with the following equations:
\begin{equation}
    p_L^{(lower)} = \sum_{\vec{w} = (0,0,...0)}^{\vec{w}_{max}} A_{\vec{w}} \, p_L^{(\vec{w})}
\end{equation}
\begin{equation}
    p_L^{(upper)} = p_L^{(lower)} + \left( 1 - \sum_{\vec{w} = (0,0,...0)}^{\vec{w}_{max}} A_{\vec{w}} \right),
\end{equation}
where $\vec{w}_{max}$ denotes the highest-weight subset that was sampled.  Notice that the lower bound assumes that all the error configurations in the unsampled subsets do not cause a logical error, while the upper bound assumes the opposite.  For low physical error rates, the two bounds essentially overlap. They start to diverge with increasing physical error rates.  In this work, we sampled the error subsets that occur with probabilities higher than $10^{-6}$ to guarantee that the separation between the lower and upper bounds is kept small.

For a more in-depth discussion of the subset sampler, the reader can consult \cite{LogicalCNOT, SteaneECBS}.  Other importance samplers have been previously applied to the simulation of QEC circuits \cite{BravyiVargo, SaschaDynamicalSubset}. 

\subsection{QEC scheme}

\begin{figure}[h!]
     \centering
     \includegraphics[width=0.85\columnwidth]{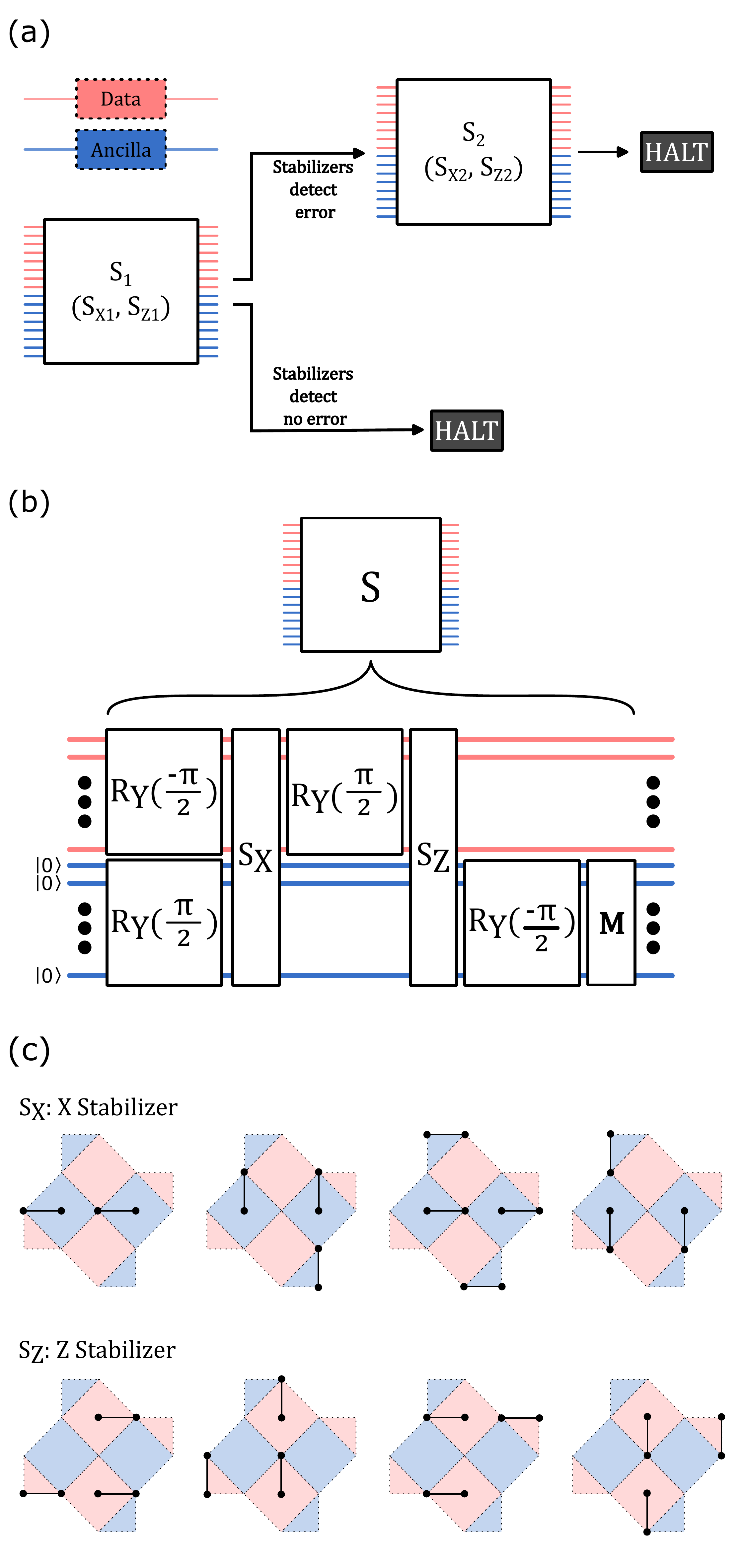}       
     \vspace{-5pt}
     \caption{ Scheme to perform FT QEC for the distance-3 codes studied in this work. (a)  The pink lines represent data qubits and the blue ones represent ancillary qubits used for the QND measurement of the stabilizers.  As explained in the main text, the data register is initialized in an error-free logical state.  $S_i$ refers to the i-th iterative measurement of all the stabilizers.  This scheme is essentially the weakly FT implementation of the adaptive protocol from \cite{AdaptiveInk} for a distance-3 code. (b) Inner structure of the procedure to measure the stabilizers of both distance-3 codes analyzed in this work. Each ancillary qubit is initialized in $|0 \rangle$.  $R_Y(\pm \pi/2)$ denotes a rotation about the Y axis by an angle $\pm \pi/2$ on each one of the qubits. These single-qubit rotations are also applied on the data qubits in order to measure the X stabilizers, since our entangling gate is CZ.  Finally, we measure all 8 ancilla qubits in the Z basis simultaneously.   (c) Each type of stabilizer (X and Z) takes 4 CZ timesteps, which are shown in the lower part of the figure. }
     \label{fig:innerstructure} 
     \vspace{-5pt}
 \end{figure}

For each QEC code studied, we simulate two fault-tolerant procedures: an error correction step and a logical $|+ \rangle_L$ state preparation (The $|0 \rangle_L$ state preparation results in essentially the same logical error rate).  For the quantum error correction step, the procedure is illustrated in Fig.~\ref{fig:innerstructure} and consists of the following steps:
\begin{enumerate}
    \item Start with a perfect (error-free) logical state on the data qubits.  The procedure is performed with $|+\rangle _L$.
    \item Using the ancillary qubits, measure all the stabilizers one time in a QND fashion.  The measurements are performed all at the end simultaneously.  These operations are depicted in Fig.~\ref{fig:innerstructure}b and \ref{fig:innerstructure}c.  
    \item If no errors are detected, the process stops. If an error is detected, measure all the stabilizers 1 more time (move to the next step).  See Fig.~\ref{fig:innerstructure}a.
    \item Use the last stabilizer outcomes to correct the logical state based on the corresponding look-up tables.  These are included in the Appendix \ref{app:Lookup}.
    \item After the correction is applied, perform noise-free error correction to project the final corrected state back to the codespace.  If the projected state is different from the initial one, a logical error has occurred.  If not, the QEC procedure has been successful.  Notice that this last step is equivalent to only counting as a logical failure the cases where the final error on the logical state is uncorrectable.  Correctable errors at the end of the procedure are not counted as failures because in the context of a FT quantum computation, they would be detected and corrected at a later stage in the circuit. 
\end{enumerate}

The quantum circuits used to perform fault-tolerant QEC on both codes are exactly the same.  This is possible because the rotated surface code corresponds to a particular gauge fixing of the square BS code.  The eigenvalues of the BS code stabilizers can be calculated by multiplying the eigenvalues of its constituent surface code stabilizers.  This is similar to the original idea by Aliferis and Cross \cite{BSAliferisCross}, with the difference that we do not measure exclusively weight-2 gauge operators, but also some weight-4 operators (the weight-4 surface code stabilizers), which consist of products of the weight-2 gauge operators.  This implies that the different performances of the two codes in the QEC step are caused solely on the classical decoding. 

\begin{figure}[h!]
     \centering
     \includegraphics[width=1.0\columnwidth]{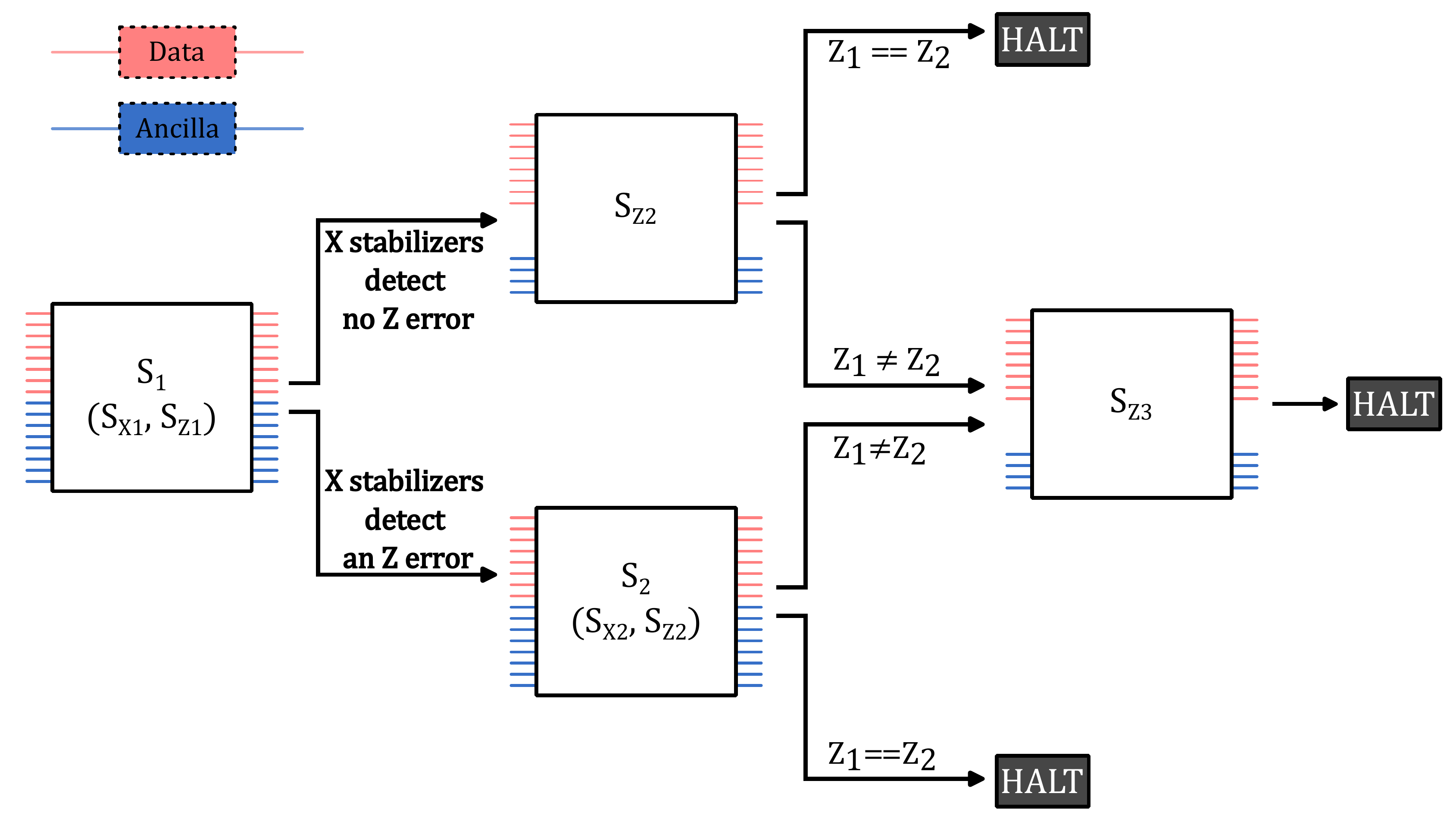}       
     \vspace{-5pt}
     \caption{Scheme to prepare the logical $|+ \rangle
     _L$ state for the surface-17 code fault-tolerantly. (a)  The pink lines represent data qubits and the blue ones represent ancillary qubits used for the QND measurement of the stabilizers.  As explained in the main text, the data register is initialized in the product state $|+\rangle ^{\otimes 9}$ and this preparation is noisy.   $S_i$ refers to the $i$-th iterative measurement of all the stabilizers.  The inner structure of each $S_i$ box is depicted in Fig.~\ref{fig:innerstructure}.}
     \label{fig:logicalstateprepscheme} 
     \vspace{-5pt}
 \end{figure}

For the fault-tolerant preparation of a logical $|+\rangle _L$ state, the procedure differs for each code.  For the surface-17 code, we employ a projective logical state preparation depicted in Fig.~\ref{fig:logicalstateprepscheme}.  Specifically, we perform the following steps:

\begin{enumerate}
    \item Prepare all data qubits in the $|+\rangle$ state, i.e., the data register is initialized in the product state $|+\rangle ^{\otimes 9}$.  This initialization is not error-free. 
    \item Using the ancillary qubits, measure all the stabilizers 1 time in a QND fashion, starting with the Z stabilizers.  The Z stabilizers have two objectives: (a) project the initial product state to one of the code subspaces and (b) detect and correct X errors.  The X stabilizers only detect and correct Z errors and are not needed for subspace projection because the initial product state is already an eigenvector of the X stabilizers.  This implies that the Z stabilizers will have to be measured at least one more time than the X stabilizers because the first time they just project the initial product state.  The initial Z-stabilizer syndrome is random and does not correspond to a real X error.
    \item If the X stabilizers detect a Z error, repeat the measurement of both the Z and X stabilizers.  Otherwise, if the X stabilizers do not detect a Z error, only repeat the measurement of the Z stabilizers.  In the latter case, we assume that no Z errors have occurred.
    \item If the first and second Z-stabilizer syndromes coincide, the procedure stops.  Otherwise, repeat the measurement of the Z stabilizers one more time and stop.  
    \item Use the final Z and X stabilizer outcomes to correct the logical state based on the corresponding look-up table.  Correct accordingly.
    \item Perform noise-free error correction to project the final corrected state back to the codespace.
\end{enumerate}

The procedure to prepare a logical $|0\rangle _L$ state is analogous to the one described above.  The difference is that the data register is initialized in $|0\rangle ^{\otimes 9}$ and the roles of the X and Z stabilizers are interchanged.

The preparation of a logical $|+\rangle_L$ state on the distance-3 BS code is considerably easier than for the surface code, provided that there is direct connectivity between the data qubits along each row, as depicted in Fig.~\ref{fig:layout_two_codes}.  For appropriate choices of the gauge qubits' states, the logical $|0\rangle_L$ and $|+\rangle_L$ states of the BS code can be expressed as products of GHZ states.  More specifically, for the distance-$3$ BS code:

\begin{align*}
 |0\rangle_L &= \frac{1}{\sqrt{2^3}} \left( \ket{+++} \, + \, \ket{---} \right) ^{\otimes 3} \quad \mbox{ along rows,} \\
  |+\rangle_L  &= \frac{1}{\sqrt{2^3}} \left( |000\rangle \, + \, |111\rangle \right) ^{\otimes 3} \quad \mbox{ along columns.}
\end{align*}
That is, a logical $|0\rangle_L$ ($|+\rangle_L$) state amounts to a product of 3 independent GHZ states along the rows (columns) of the code. Fig.~\ref{fig:logstateprepBS} illustrates the circuit used to prepare the GHZ state $\frac{1}{\sqrt{2}} (|000\rangle + |111\rangle)$. 
  
\begin{figure}[h]
    \[ 
    \Qcircuit @C=0.2em @R=.2em @! {
    \lstick{\ket{0}} & \gate{R_Y(\pi/2)} & \ctrl{1} & \qw & \gate{R_Y(-\pi/2)} \\
    \lstick{\ket{0}} & \gate{R_Y(\pi/2)} & \ctrl{-1} & \ctrl{1} & \qw \\
    \lstick{\ket{0}} & \gate{R_Y(\pi/2)} & \qw & \ctrl{-1} & \gate{R_Y(-\pi/2)} \\
    }\]
        
    \caption{Circuit used in our simulations to prepare the GHZ state $\frac{1}{\sqrt{2}} (|000\rangle + |111\rangle)$.  $R_Y(\pm \pi/2)$ denotes a rotation about the Y axis by $\pm \pi/2$.  The vertical lines denote CZ entangling gates.}
    \label{fig:logstateprepBS}
    \vspace{-5pt}
\end{figure}

The circuit depicted in Fig.~\ref{fig:logstateprepBS} is naturally FT for a distance-$3$ code because when a single-qubit error propagates through an entangling gate the resulting $2$-qubit error is equivalent to another single-qubit error up to one of the GHZ stabilizers.  This implies that for the distance-$3$ BS code the state preparation of the logical $|0\rangle_L$ and $|+\rangle_L$ states can be fully coherent (not projective via stabilizer measurements) and still be FT, as has been already experimentally demonstrated in trapped ions \cite{Egan2021}.  For distances $d>3$, the circuits are not naturally FT and flag-like qubits are needed to verify the GHZ states \cite{SteaneECBS}.

Since our proposed layout only has direct qubit connectivity between qubits along the same rows, to prepare $|+\rangle_L$, we first implement the circuit in Fig.~\ref{fig:logstateprepBS} along each row of the code and then relabel the qubits to exchange the rows and columns.  This is an entirely classical procedure that can be done offline and therefore results in no errors.  The same relabeling is performed when implementing a logical H gate on both the surface and the BS square codes.

\section{Results}  \label{sec:Results}

Table \ref{tab:parameters} summarizes the state-of-the-art values of key parameters for silicon-based spin qubits (specifically those within Si/SiGe quantum wells). These parameters are the starting point of our simulations throughout this work. For a broader analysis of semiconductor spin qubit performance metrics, please refer to  \cite{stano2021review}. 

\begin{figure}
    \centering
    \includegraphics[width=1 \columnwidth]{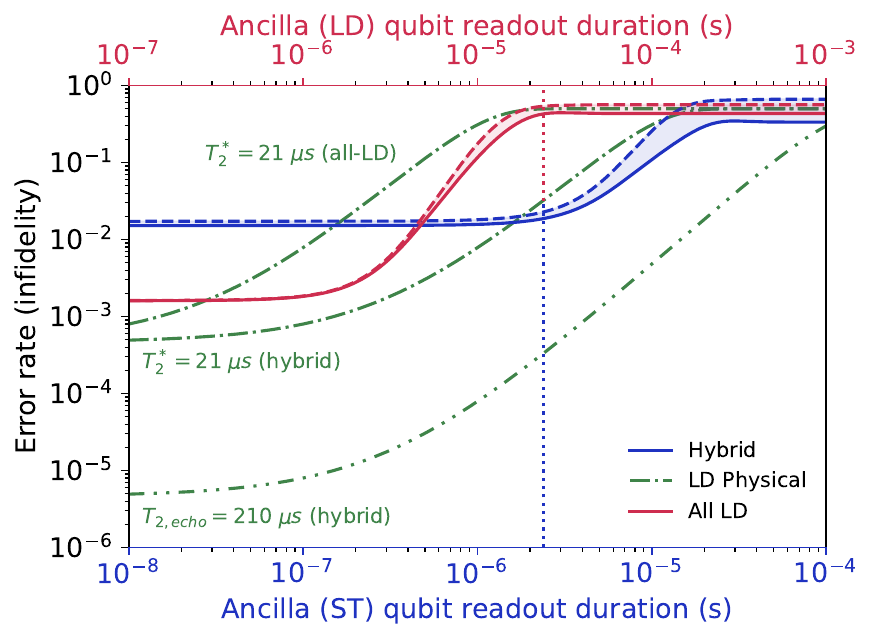}
       \vspace{0pt}
    \caption{Logical error rate for 1 QEC step of the surface-17 code for two different qubit encoding schemes as a function of the readout duration of the ancillary qubits.  The readout duration of the LD (ST) qubit appears in the upper (lower) $x$ axis.  We have constructed the upper and lower $x$ axes such that the current ratio between the readout duration of the LD qubit to the ST qubit ($10$) is maintained.  The vertical line indicates the state-of-the-art values of the readout durations for each qubit.  The green curves correspond to the infidelity of a physical qubit that starts in the $|+ \rangle$ and undergoes dephasing for the duration of the whole QEC step of the all-LD or hybrid scheme, with a coherence time of 21 $\mu$s or 210 $\mu$s achievable via spin echo. For long readout times, we observe a separation between the lower (solid lines) and upper (dashed lines) bounds of the logical error rates, which occurs because the probability of occurrence of the unsampled error subsets becomes considerable.  For the state-of-the-art ST readout time ($2.4 \, \mu s$), the hybrid logical qubit results in a slightly lower error rate than the LD physical qubit.}   
   \label{fig:surfaceQEC}
\end{figure}

 Figure \ref{fig:surfaceQEC} shows the logical error rate for one QEC step of the surface-17 code with the two different encoding schemes.  The physical error rates employed for this simulation are presented in Table \ref{tab:parameters}, except for the qubit readout duration, which is taken as the independent variable.  The state-of-the-art readout duration of the LD (ST) qubit is $24 \ \mu$s ($2.4 \ \mu$s).  In order to compare the all-LD scheme and the hybrid scheme, we have assumed that the ratio between the two readout durations will be kept constant.  Therefore, Fig.~\ref{fig:surfaceQEC} sets the upper X axis to be exactly $10$ times the lower X axis.  The vertical dotted line marks the state-of-the-art value.      
\begin{table}
\centering
\begin{tabular}{ | c | c | c | c |}
\hline Parameter &  \; LD qubit \; & ST qubit \;    \\ \hline   
$T_2^*$ ($\mu$s) \strut & $ 21.0 $ \cite{struck2020low} & {$ 14.8^\dagger $ }\\ \hline  
1-q gate infidelity  \strut & \,$4 \, \times 10^{-4}$\cite{philips2022universal,mills2022high} & $4\, \times 10^{-3}$ \cite{takeda2020resonantly}  \\ \hline  
2-q gate time (ns) \strut & $40 $ (CZ)  \cite{mills2022two} & $40$$^{\dagger\dagger}$ \\ \hline  
2-q gate infidelity \strut & \,$2  \times 10^{-3}$ \cite{mills2022two} & $4\, \times 10^{-3}$  $^{\ddagger}$ \\ \hline  
Readout time ($\mu$s)\strut & $24$ \cite{volk2019fast} & $2.4$ \cite{takeda2024rapid}  \\ \hline  
Readout infidelity  \strut  & $2.4 \times 10^{-3}$ \cite{mills2022high}&  $4 \times 10^{-4}$ \cite{takeda2024rapid} $^{\ddagger\ddagger}$  \\\hline  
State prep. infidelity \strut  &  $6.5 \times 10^{-3}$ \cite{mills2022high}  & $4 \times 10^{-3}$ \cite{niegemann2022parity}  \\ \hline  
Total QEC cycle ($\mu$s) \strut  &  $24.32 \, - \, 48.64$   & $2.72 \, - \, 5.44$  \\ \hline
\end{tabular}
\caption{Qubit parameters used for the simulations. $^\dagger$The coherence time of the ST qubit is calculated as that of the LD qubit divided by $\sqrt{2}$, which arises when $\Delta E_z\gg J_{ST}$, and assuming uncorrelated fluctuations of the Zeeman splittings of the two spins. $^{\dagger\dagger}$The physical mechanism of the two-qubit gate for the ST qubit is the same as that for the LD qubit, so the gate time is assumed to be equal. $^{\ddagger}$In this table and throughout this paper, the ST 2-qubit gate refers to a CZ gate between an LD qubit and an ST qubit. Two-qubit operations have been performed between an LD and an ST qubit in GaAs \cite{noiri2018fast}; these are expected to improve by more than ten-fold when the same operations are implemented in isotopically purified silicon. In particular, the infidelity assumed for this gate reflects the reduced ST coherence time, leading to an approximately factor-of-two increase relative to the LD–LD CZ gate.  $^{\ddagger\ddagger}$The readout infidelity related to the integration time after ramping to the readout point; see the main text for further discussion. The total QEC duration time is calculated with Equation \ref{eq:QECduration}.  The lower (upper) limit corresponds to $1$ ($2$) round(s) of stabilizer measurements.}

\label{tab:parameters}
\end{table}

The first observation that can be extracted from Fig.~\ref{fig:surfaceQEC} is that, for current physical error rates, the hybrid scheme outperforms the all-LD scheme by more than an order of magnitude when looking at the state-of-the-art readout times (vertical dotted line).  The LD qubit readout duration is so long relative to the lifetime of the physical qubits that, after a QEC step, the all-LD logical qubit has essentially dephased to a fully mixed state.   The great advantage of the hybrid scheme thus originates from the reduced dephasing because of the shorter readout duration of the ST qubit.

In the limit of very short readout durations, the all-LD logical qubit reaches a saturation error rate of $1.6 \times 10^{-3}$, compared to  $1.6 \times 10^{-2}$ for the hybrid scheme.  This outperformance originates in the higher fidelity of the LD qubit $1$- and $2$-qubit gates.  Nevertheless, this assumes that readout durations can be shortened arbitrarily and without causing any effect on the other parameters.  In a more realistic scenario, there is a trade-off between the readout duration and its fidelity.     

For the current fidelity and duration values, the hybrid logical scheme even outperforms a physical LD qubit ($T^*_2 = 21\ \mu$s) initialized in the $|+ \rangle$ state and subject to dephasing (Z) noise during the duration of the whole QEC process for both the hybrid  and the all-LD schemes (green dashed-dotted lines in Fig.~\ref{fig:surfaceQEC}).  Since the QEC process follows an adaptive procedure (see Fig.~\ref{fig:innerstructure}), for a given run, its duration is not determined \textit{a priori}.  More specifically, the QEC process could stop after either one or two rounds of stabilizer measurements.  The green curves on Fig.~\ref{fig:surfaceQEC} correspond to the maximal duration, where the QEC process involves two rounds of stabilizer measurements.  For each scheme, the total durations are presented in Table \ref{tab:parameters}.

To calculate the duration of $1$ round of stabilizer measurements, we employ the following equation:
\begin{equation}
    t_{\textrm{round}} = 8 \, t_{\textrm{CZ}} + t_{\textrm{readout}}.
\label{eq:QECduration}
\end{equation}
However, a simple spin echo pulse on the physical qubit is enough to turn it into a quantum memory superior to the logical qubit, as depicted by the green curve labeled by $T_{2,\textrm{echo}}$ in Fig.~\ref{fig:surfaceQEC}.

\subsection{Effect of the tradeoff between the readout integration time and its fidelity}

Having established the superior performance of the hybrid encoding scheme over the all-LD scheme under current experimental conditions, we now focus on the hybrid scheme and analyze the trade-off between readout integration time and readout fidelity for the ST qubit. We assume a spin readout based on PSB, as demonstrated in Ref.~\cite{takeda2024rapid}.

The total readout time of an ST qubit consists of two components: a \textit{ramping time}, during which the system is pulsed to the measurement point in gate-voltage space, and an \textit{integration time}, over which the charge sensor signal is collected and averaged to distinguish between singlet and triplet states. This is expressed as:
\begin{equation}
t_{\textrm{readout}} = t_{\textrm{ramp}} + t_{\textrm{int}}.
\end{equation}

For ST qubits, there is a clear relationship between the integration time and the readout infidelity, as shown in Fig.~\ref{fig:ST_readout_infidelity}.  Up to about $2 \, \mu s$, the readout infidelity decreases with increasing integration time, by improving the signal-to-noise ratio.  For longer integration times ($t_\textrm{int} > 2 \, \mu s$), the readout infidelity starts to increase because of the relaxation of the triplets.  Apart from this, the longer the integration time, the longer the idling time of the data qubits, and, consequently, the more intense their dephasing. In our simulations, we fix the ramping time to $t_\textrm{ramp} = 0.4 \, \mu s$ and vary the integration time $t_\textrm{int}$ to explore this trade-off.

\begin{figure}
    \centering
    \includegraphics[width=1 \columnwidth]{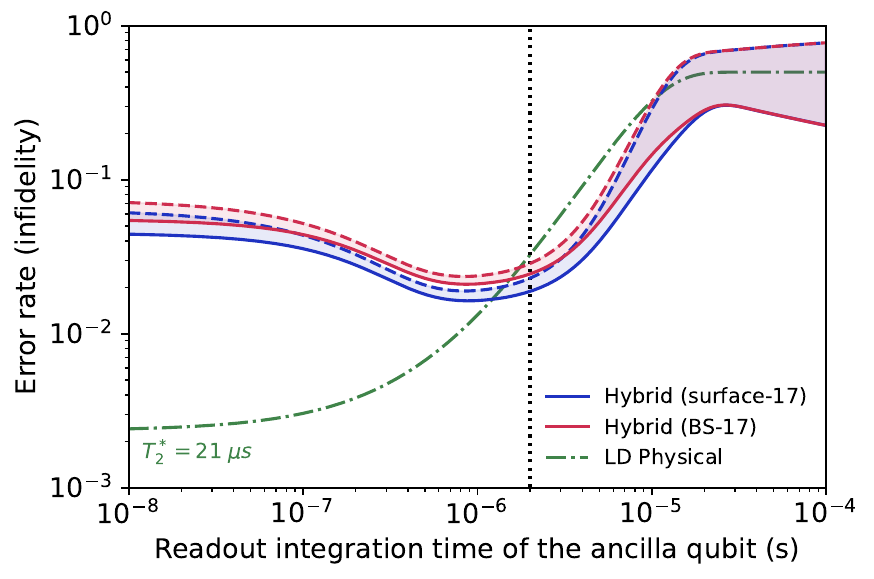}
       \vspace{0pt}
    \caption{Logical error rate for $1$ QEC step of the surface-$17$ and the BS-$17$ codes for the hybrid encoding scheme as a function of the readout integration time of the ancillary qubits.  The ramping time is kept constant at $0.4~\mu$s.  As in Fig.~\ref{fig:surfaceQEC}, the green curve correspond to the infidelity of a physical qubit that starts in the $|+ \rangle$ and undergoes dephasing for the duration of the whole QEC step of the hybrid scheme.  The vertical dotted line marks the state-of-the-art integration time of $2.0~\mu$s.}   
   \label{fig:BSvssurfaceQEC}
\end{figure}

Remarkably, after including the effect of the integration time on the readout infidelity, the hybrid logical qubit still slightly outperforms a physical qubit at the state-of-the-art readout duration, as shown in Fig.~\ref{fig:BSvssurfaceQEC}.  For this integration time ($2.0\ \mu$s), the surface-$17$ code results in a logical error rate of $(2.1 \pm 0.2) \times 10^{-2}$, while the BS-$17$ code results in a logical error rate of $(2.7 \pm 0.2) \times 10^{-2}$.  Although the lowest logical error rates are achieved at an integration time of around $0.88 \ \mu$s, at $2.0 \ \mu$s the resulting logical error rates are close to the minimum value.  Notably, this integration time resulting in the lowest logical error rate is lower than the integration time that results in the lowest readout infidelity (around $1.4 \ \mu$s).  Het\'enyi and Wootton have also observed that, for ST qubits, the readout integration time resulting in the lowest logical error rate is lower than the readout integration time resulting in the lowest readout infidelity \cite{hetenyi2023tailoring}.  The reason is essentially the same as in our case: {even though a shorter integration readout time increases the readout infidelity because of a lower signal-to-noise ratio, this is compensated by a much lower dephasing rate on the idle data qubits because of the shorter waiting time during stabilizer measurements}.   Estimating the optimal integration time will be crucial when running actual QEC experimental demonstrations on QD architectures.  For QEC codes of higher distance, for which the limiting factor will most likely be the memory errors, it might be possible to employ soft decoding of the readout signal to shorten the optimal readout integration time even further.

For very short integration times, the logical qubit no longer outperforms the physical qubit.  As before, a simple spin echo pulse on the physical qubit is enough to turn it into a quantum memory superior to the logical qubit for all integration times.

The performance of the BS-$17$ code follows exactly the same trend as the surface-$17$ code, but with a slightly larger logical error rate.  This is a result exclusively of the decoding process.  There are various ways in which one can fault-tolerantly measure the stabilizers of a subsystem code like the BS-$17$ code.  On one extreme, it is possible to measure the stabilizers directly using a single ancillary qubit \cite{BSBareAncilla, Egan2021, SteaneECBS}, which is convenient in systems with long-range qubit connectivity.  On the other extreme, for systems with only nearest-neighbor connectivity, it is also possible to infer the outcome of the high-weight stabilizers by measuring its constituent weight-$2$ gauge operators and simply calculating the total parity \cite{BSAliferisCross}.  

We take an intermediate approach and measure some weight-$2$ and some weight-$4$ operators and then calculate the total parity.  In fact, we measure the surface-$17$ stabilizers and the quantum circuit that we implement is exactly the same for both codes.  The only difference lies in the way in which we interpret the outcomes of these operators.  In the surface-$17$ code, we use all the outcomes to infer the error.  In the BS-$17$, we interpret these operators not as stabilizers, but rather as gauge operators, and multiply them to obtain the total parity of the high-weight stabilizers.  Since the quantum circuits for both codes are exactly the same, it is this discarding of classical information what causes the slightly worse performance of the BS-$17$ code.

\subsection{Impact of each physical parameter on the logical error rate}

\begin{figure*}[t]
    \centering
    \includegraphics[width=1 \textwidth]{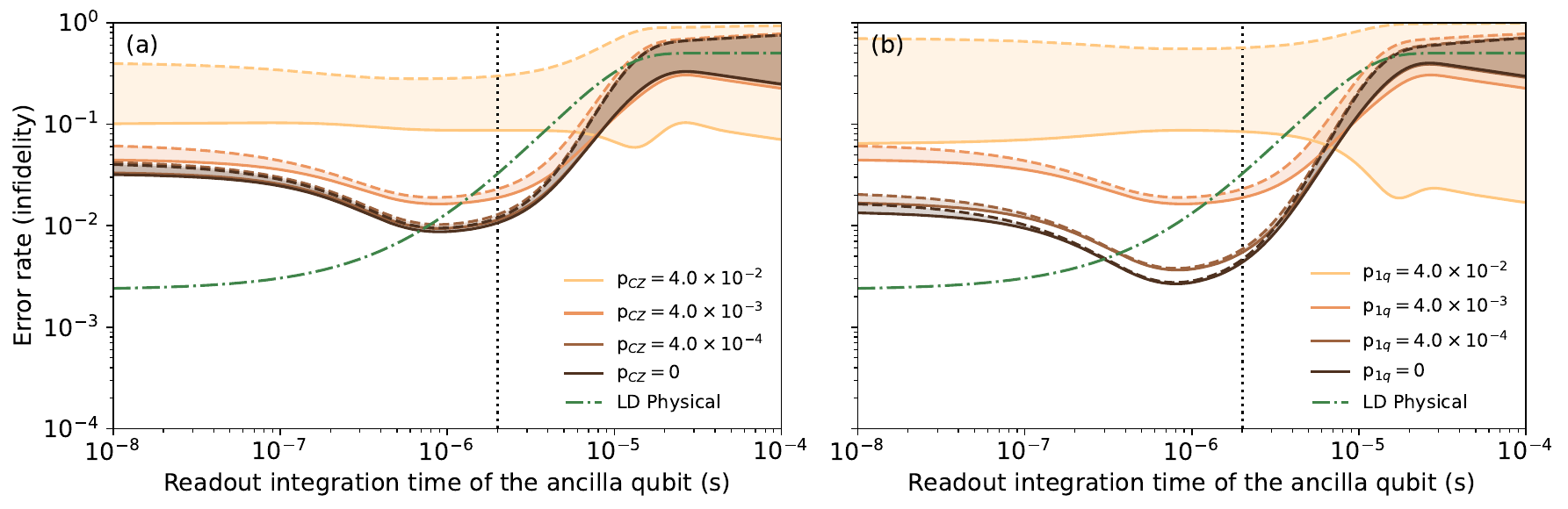} 
       \vspace{0pt}
    \caption{Logical error rate for 1 QEC step of the surface-17 in the hybrid encoding scheme as a function of the readout integration time of the ancillary qubits, for various (a) CZ error probabilities and (d) single-qubit gate error probabilities.  The ramping time is also kept constant at $0.4~\mu $s.  In both panels, the dashed-dotted green curves correspond to the infidelity of a physical qubit with $T_2^*=21~\mu$s that starts in the $|+ \rangle$ state and undergoes dephasing for the duration of the whole QEC step of the hybrid scheme.  The vertical dotted line marks the state-of-the-art integration time of $2.0~\mu$s.}   
    \label{fig:2params}
\end{figure*}

We now explore the impact of several physical parameters on the logical error rate, with the goal of guiding future experimental work.  For 1 FT QEC step (involving $1$ or $2$ rounds of stabilizer measurements) of the surface-$17$ code implemented on the hybrid encoding scheme, we explore the impact of the two-qubit gate error rate and duration, the state preparation infidelity, the single-qubit gate infidelity, and the $T^{*}_2$ time of the physical qubits.  The results for the BS-$17$ code are essentially the same.  

Figure \ref{fig:2params} shows the impact of the 2 physical parameters most determinant on the logical error rate of the surface-17 code with the hybrid scheme, namely the two-qubit and single-qubit gates' infidelities.  In each panel, the darkest curve marks the lower bound on the logical error rate for each particular physical error parameter.    Since the two-qubit gate duration and the state-preparation infidelity both have a negligible effect on the logical error rate, we leave their discussion to the Appendix \ref{app:othereffects}.   

\textbf{Infidelity of the CZ gate:} The first parameter to analyze is the error rate of the CZ gate, the entangling gate of choice for spin qubits. As shown in Fig.~\ref{fig:2params}a, reducing the CZ gate infidelity from $4.0 \times 10^{-3}$ to $4.0 \times 10^{-4}$ leads to a substantial improvement in the logical error rate. For the higher infidelity, the logical error rate reaches a minimum of $(1.7 \pm 0.1) \times 10^{-2}$, whereas with the lower infidelity, the minimum improves to $(0.97 \pm 0.03) \times 10^{-2}$, both evaluated at a readout integration time of $0.88 \ \mu\text{s}$..  That is, the minimal logical error rate would be reduced $1.75$ times by decreasing the CZ infidelity by one order of magnitude.  At this point, the curve already almost saturates the lower bound (darkest curve), so further reductions in the CZ infidelity would not have any impact if other error rates are not reduced concurrently.

\textbf{Infidelity of single-qubit gates:}  As seen in Fig.~\ref{fig:2params}b, decreasing the infidelity of the single-qubit gates from the current value by $10$ times has, of all parameters, the largest effect on the logical error rate.  The minimal logical error rate of $(1.7 \pm 0.1) \times 10^{-2}$ (at $t_{\textrm{int}} = 0.88 \ \mu$s) would be reduced to $(3.7 \pm 0.1) \times 10^{-3}$ (at $t_{\textrm{int}} = 0.82 \ \mu$s). This is a reduction of $4.6$ times by decreasing $p_{\textrm{1q}}$ by a factor of $10$.  At this point, the lower bound is already almost saturated, so further reductions in the single-qubit gate infidelity would not have a significant impact.

What causes this high sensitivity of the logical error rate on this parameter?  The only single-qubit gates employed in our QEC circuit are $\pm \pi/2$ rotations about the $Y$ axis, with associated noise in the form of stochastic $Y$ errors.  These single-qubit gates are used in two different contexts in the circuit:

(1) When measuring $X$ stabilizers, on the data qubits, a $R_Y(\mp \pi/2)$ before and after the CZ gates.  The $Y$ errors that might occur after these rotations are eventually detected by subsequent stabilizer measurements, but they do constitute a considerable source of noise on the data qubits.  This suggests that it is probably recommendable to measure first the $X$ stabilizers, in order to detect these $Y$ errors when measuring the $Z$ stabilizers.  

(2) When measuring both types of stabilizers, on the ancillary qubits, a $R_Y(\pm \pi/2)$ rotation immediately after (before) the state preparation (readout) to transform between the $Z$ and $X$ bases.  The $Y$ errors that occur after the $R_Y(+ \pi/2)$ after the state preparation will propagate to the data qubits, but will not generate an actual error, since the propagated operator is actually equivalent to the stabilizer being measured.  However, $Y$ errors after these $Y$ rotations will produce measurement errors.  Since there are two locations on each ancillary qubit where these $Y$ errors can occur, this increases the effective readout error to:
\begin{equation}
    p_{\textrm{readout,eff}} = p_{\textrm{readout}} + 2 \, p_{1q} + O(p^2).     
\end{equation}
 
The single-qubit gate errors can have a considerable effect on the readout error rate.  However, as explained in Appendix \ref{app:othereffects}, this effective readout error does not have a considerable effect on the logical error rate.  Therefore, the significant effect of the single-qubit gate errors on the logical state originates on the $R_Y(\mp \pi/2)$ gates on the data qubits.

\begin{figure*}
    \centering
    \includegraphics[width=1 \textwidth]{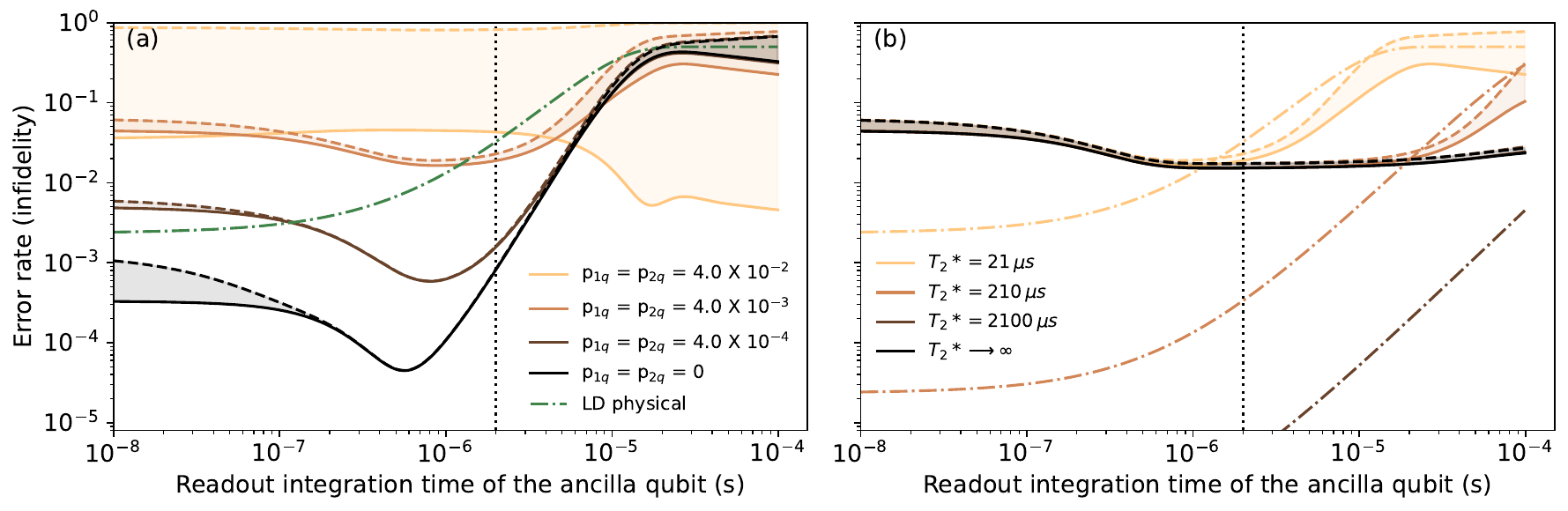}
       \vspace{0pt}
    \caption{Logical error rate for 1 QEC step of the surface-17 code with the hybrid encoding scheme as a function of the readout integration time of the ancillary qubits, for (a) the simultaneous reduction of $p_{1q}$ and $p_{2q}$ and (b) various $T_2^*$ values.  In (a), by reducing $p_{1q}$ and $p_{2q}$ simultaneously, a considerably lower logical error rate is achieved.  In (b), the $T_2^*$ values correspond to the LD qubits', and for the ST qubits are taken to be exactly $1/\sqrt{2}$ that of the LD qubits'.  The dashed-dotted curves correspond to the infidelity of a physical qubit that starts in the $|+ \rangle$ state and undergoes dephasing for the duration of the whole QEC step of the hybrid scheme.   The vertical dotted line marks the state-of-the-art integration time of $2.0~\mu$s.}   
   \label{fig:collection10-11}
\end{figure*}


\textbf{Simultaneous reduction of $p_{1q}$ and $p_{2q}$:} As shown in Fig.~\ref{fig:collection10-11}a, a simultaneous reduction of the single-qubit and two-qubit gates error rates would have the most drastic effect on the logical error rate.  Decreasing $p_{1q}$ and $p_{2q}$ by $10$ times would lead to a minimum logical error rate below $10^{-3}$.  In turn, this would still be far above the lower bound obtained for $p_{1q} = p_{2q} = 0$, which would be less than $10^{-4}$.  This implies that for experimental demonstrations of these distance-3 codes, a concurrent reduction in the error rates of 1- and 2-qubit gates would have the largest impact on the logical error rate.


\textbf{$T_2^*$ time:} In order to distinguish between the effects of memory errors and gate errors, we simulate the QEC circuit for various $T_2^*$ values and show the results in Fig.~\ref{fig:collection10-11}b.  The lower (upper) bound of the logical errors rates are displayed as solid (dashed) curves.  The physical error rates correspond to the dashed-dotted curves, with the ones for $T_2^* = 21 \mu$s being the same as shown in previous plots.  There are several interesting trends. First, as the $T_2^*$ time increases, the infidelities of the physical and logical qubits both decrease, as expected.  Second, the break-even integration time above which the logical infidelity is lower than the physical infidelity increases with increasing $T_2^*$ times, also as one would expect.  Finally, in the limit of $T_2^* \to \infty$, we obtain the lower bound where we only see the effect of gate errors.  Only the logical infidelity is shown, since the physical infidelity would be equal to $0$ for all readout integration times.  The value of this lower bound is given by all the gate errors (single-qubit and 2-qubit gates, state preparation, and measurements which, following what was shown in Fig.~\ref{fig:2params}, are dominated by single-qubit and 2-qubit gates).  Since the readout infidelity increases for short readout integration times, this causes a considerable increase in the logical error rate.  

The most remarkable feature is that at the state-of-the-art readout integration time and below, the logical error rate curve for $T_2^* = 21 \ \mu$s practically saturates the lower bound ($T_2^* \to \infty$).  This means that \textit{the logical error rate is fundamentally limited by gate errors and not memory errors}.  This is a rather unexpected result that probably only holds for low-distance codes.  In our case, the distance-3 QEC protocol is fast, since we only do one or at most two repetitions of the stabilizer measurements.  High-distance codes will require more repetitions ($\sim d$ for codes that can be decoded with a matching algorithm) and memory errors will most likely have a significant effect on the logical fidelity.  Nevertheless, for proof-of-concept experiments involving distance-3 codes, our results show that gate errors will have a much larger impact than memory errors, when using the hybrid LD-ST scheme.     

\subsection{Logical state preparation: key advantage of the BS-17 over the surface-17 code}

\begin{figure}
    \centering
    \includegraphics[width=1 \columnwidth]{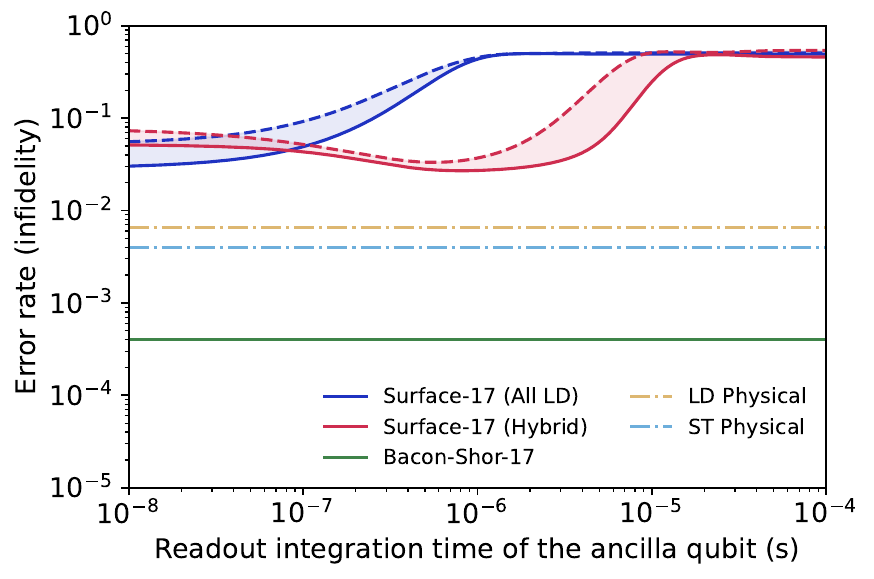} 
       \vspace{0pt}
    \caption{Logical error rate for the FT logical state preparation of the surface-17 code and the BS-17 code a function of the readout integration time of the ancillary qubits. For the physical qubits, we plot the current state-preparation infidelities, which do not depend on the readout integration time.  The logical error rate for the state preparation of the BS-17 also does not depend on the readout integration time, since the circuit is completely unitary and does not rely on projective ancillary measurements.  For the surface-17 code with the hybrid encoding scheme, we include the integration time effect on the readout infidelity.  For the the surface-17 code with the all-LD scheme, we assume $t_{\textrm{readout, LD}} = 10 \, t_{\textrm{readout, ST}}$.}   
   \label{fig:FTstateprep}
\end{figure}

Figure \ref{fig:FTstateprep} shows the error rate for the FT logical state preparation of a $|+ \rangle _L$ state for the surface-17 and the BS-17 codes.  (The results for the preparation of $|0 \rangle _L$ are essentially the same.)  The most salient result is that the BS-17 code outperforms the rest by a wide margin, with a logical error rate of $4.05 \times 10^{-4}$.  It beats the surface-17 code with both the all-LD and hybrid encodings, and even the physical state preparations by about 1 order of magnitude.  This happens because the logical state preparation of $|0 \rangle _L$ and $|+ \rangle _L$ states in the BS-17 code is FT for the unitary circuit and it simply amounts to preparing three separate GHZ states, as shown in Fig.~\ref{fig:logstateprepBS}. Since these logical state preparations on the BS-17 code do not rely on measurements, the logical error rate is just a horizontal line as a function of the readout integration time.  This constitutes the key advantage of the BS-17 code over the surface-17 code, which needs projective state preparation to be made FT.  Crucially, this advantage of the BS-17 code requires a qubit configuration where the data qubits along the rows (or alternatively the columns) of the code are connected, as depicted in Fig.~\ref{fig:layout_two_codes}, in order to be able to directly apply entangling gates between them.    

Notice that we only need connectivity between the data qubits \textit{either} along the rows \textit{or} columns of the code to be able to prepare \textit{both} the $|0 \rangle _L$ and $|+ \rangle _L$ states of the BS-17 in a unitary fashion.  Even though $|0 \rangle _L \propto (\ket{+++} + \ket{---})^{\otimes 3}$ is defined along the rows and $|+ \rangle _L \propto (|000 \rangle + |111 \rangle)^{\otimes 3}$ is defined along the columns, we can simply prepare both states along the rows and then relabel the qubits offline if the desired state was to be defined along the columns, in a similar spirit as how the transversal logical Hadamard gate requires a qubit relabeling in both the surface and BS square codes. 

For the surface-17 code, neither the hybrid nor the all-LD encoding schemes manage to outperform the physical qubits for any readout integration times.  However, the hybrid scheme clearly has superior performance.

\section{Conclusions and Outlook} \label{sec:Conclusions}

In this paper, we have explored two alternative encoding schemes to implement a distance-$3$ error-corrected logical qubit on a QD, spin-qubit system: a scheme in which all the qubits are LD qubits (all-LD) and another one where the data qubits are LD while the ancilla qubits are ST (hybrid).  We have also studied two different compass codes: the surface-$17$ and the BS-$17$.   For state-of-the-art parameters, the hybrid scheme outperforms the all-LD scheme by more than one order of magnitude.  The hybrid scheme's superior performance is essentially due to the much shorter readout duration, which drastically reduces the dephasing on the data qubits during the measurements of the ancilla qubits.  The readout duration of the LD qubits is so long that the all-LD logical qubit is fully dephased after a FT QEC step.

In the hybrid scheme, the ST readout duration is so short compared to the physical qubits' lifetimes, that the logical error rate is completely dominated by gate errors and not memory errors.  Although memory errors will most likely become significant for codes of higher distance (since more stabilizer measurement repetitions are necessary to guarantee FT), this surprising result will be important for future proof-of-concept experimental demonstrations of distance-$3$ codes on spin qubits.  Of all possible gate errors, the single-qubit and two-qubit gate errors have the greatest effect on the logical error rate.  We have also found the optimal integration times that result in the lowest logical error rates.  The current $2.0 \, \mu s$ ST qubit integration time results in a logical error rate that is fairly close to the minimum.

For the hybrid scheme, we have compared the surface-$17$ and the BS-$17$ codes. We have found that the surface-$17$ code performs slightly better on the FT QEC.  However, the BS-$17$ code outperforms the surface-$17$ by $2$ orders of magnitude on the FT logical state preparation of $|0\rangle_L$ and $|+\rangle_L$.  This extremely low logical error rate is obtained because the distance-$3$ BS-$17$ code supports a very shallow and fully coherent (as opposed to projective) preparation of these two logical states.  

Several questions remained to be answered.  We have included several sources of noise in our simulations.  However, we have assumed that noise is spatially uncorrelated.  Several experimental studies  have revealed that spin qubits exhibit both spatial and temporal noise correlations \cite{yoneda2023noise, rojas2023spatial, rojas2026origins, rojas2026inferring}.  Understanding the effect of these correlations on the performance of QEC codes will be crucial.  In future work we also plan to go beyond distance-$3$ codes.  To construct an algorithmically useful logical qubit, logical error rates would have to be extremely low.  At this point, memory errors will play a very significant role and codes tailored to the biased noise present in spin qubits will likely be crucial.  

Finally, recent experimental demonstrations of electron shuttling in quantum dots \cite{ShuttlingLieven,ShuttlingLars} have opened up a vast repertoire of QEC codes and protocols that are inaccessible to architectures with only nearest-neighbor interactions.  In particular, it would be very interesting to explore how electron shuttling can be employed to implement high-rate, low-overhead quantum low-density parity-check (LDPC) codes \cite{LDPC_IBM,LDPC_China,TileCodes} and concatenated code schemes \cite{HayataConcatenate} on spin qubits.  Furthermore, electron shuttling will be crucial in implementing standard 2-D QEC codes (like the ones analyzed in this work) on QD layouts that are effectively ``1.5-D'', \textit{i.e.}, layouts that can extend indefinitely along one direction but have a fixed length along the perpendicular direction, like the $2 \times N$ \cite{2XN} and $3 \times N$ \cite{3XN} QD arrays reported recently.

\section{Acknowledgments}

This research was financially supported by the Top Research Centers in Taiwan Key Fields Program, funded by the Ministry of Education (MOE), Taiwan.  M.G. acknowledges additional support by the Research Office of the University of Costa Rica.   Some of the simulations were run on the computer cluster of the Research Center for Materials Science and Engineering (CICIMA) of the University of Costa Rica.  M.G. thanks Federico Mu\~noz for his help in setting up the simulations on the computer cluster.   

\section{Data availability}

The simulation toolkit \cite{GithubRepo} and the complete results dataset \cite{ZenodoRepo} are publicly available online.

\appendix

\section{Lookup tables} \label{app:Lookup}

For the space decoding of the QEC codes we used lookup tables.  X and Z errors were corrected independently. The lookup tables of the distance-3 rotated surface code are shown in Tables \ref{table:surface17Xstabs} and \ref{table:surface17Zstabs} for the X and Z stabilizers, respectively. 

\begin{table}
\centering
\vspace{10pt}
\begin{tabular}{ | c  c  c  c | c |}
\hline \; $S_X^{(1)}$ \; & \; $S_X^{(2)}$ \; & \; $S_X^{(3)}$ \; & \; $S_X^{(4)}$ \; & \; Error \;   \\ \hline \hline  
$+1$ &  $+1$ & $+1$ & $+1$ & $I$  \\   \hline  
$+1$ &  $+1$ & $+1$ & $-1$ & $Z_6$  \\   \hline  
$+1$ &  $+1$ & $-1$ & $+1$ & $Z_5$  \\   \hline  
$+1$ &  $+1$ & $-1$ & $-1$ & $Z_7$  \\   \hline  
$+1$ &  $-1$ & $+1$ & $+1$ & $Z_0$  \\   \hline  
$+1$ &  $-1$ & $+1$ & $-1$ & $Z_3Z_6$  \\   \hline  
$+1$ &  $-1$ & $-1$ & $+1$ & $Z_4$  \\   \hline  
$+1$ &  $-1$ & $-1$ & $-1$ & $Z_4Z_6$  \\   \hline  
$-1$ &  $+1$ & $+1$ & $+1$ & $Z_2$  \\   \hline  
$-1$ &  $+1$ & $+1$ & $-1$ & $Z_2Z_6$  \\   \hline  
$-1$ &  $+1$ & $-1$ & $+1$ & $Z_2Z_5$  \\   \hline  
$-1$ &  $+1$ & $-1$ & $-1$ & $Z_2Z_7$  \\   \hline  
$-1$ &  $-1$ & $+1$ & $+1$ & $Z_1$  \\   \hline  
$-1$ &  $-1$ & $+1$ & $-1$ & $Z_1Z_6$  \\   \hline  
$-1$ &  $-1$ & $-1$ & $+1$ & $Z_2Z_4$  \\   \hline  
$-1$ &  $-1$ & $-1$ & $-1$ & $Z_1Z_7$  \\   \hline  
\end{tabular}
\vspace{10pt}
\caption{ Lookup-table for the X stabilizers of the surface-17 code.  }
\vspace{10pt}
\label{table:surface17Xstabs}
\end{table}

\begin{table}
\centering
\vspace{10pt}
\begin{tabular}{ | c  c  c  c | c |}
\hline \; $S_Z^{(1)}$ \; & \; $S_Z^{(2)}$ \; & \; $S_Z^{(3)}$ \; & \; $S_Z^{(4)}$ \; & \; Error \;   \\ \hline \hline  
$+1$ &  $+1$ & $+1$ & $+1$ & $I$  \\   \hline  
$+1$ &  $+1$ & $+1$ & $-1$ & $X_8$  \\   \hline  
$+1$ &  $+1$ & $-1$ & $+1$ & $X_6$  \\   \hline  
$+1$ &  $+1$ & $-1$ & $-1$ & $X_7X_8$  \\   \hline  
$+1$ &  $-1$ & $+1$ & $+1$ & $X_1$  \\   \hline  
$+1$ &  $-1$ & $+1$ & $-1$ & $X_5$  \\   \hline  
$+1$ &  $-1$ & $-1$ & $+1$ & $X_4$  \\   \hline  
$+1$ &  $-1$ & $-1$ & $-1$ & $X_4X_8$  \\   \hline  
$-1$ &  $+1$ & $+1$ & $+1$ & $X_0$  \\   \hline  
$-1$ &  $+1$ & $+1$ & $-1$ & $X_0X_8$  \\   \hline  
$-1$ &  $+1$ & $-1$ & $+1$ & $X_3$  \\   \hline  
$-1$ &  $+1$ & $-1$ & $-1$ & $X_3X_8$  \\   \hline  
$-1$ &  $-1$ & $+1$ & $+1$ & $X_0X_1$  \\   \hline  
$-1$ &  $-1$ & $+1$ & $-1$ & $X_0X_5$  \\   \hline  
$-1$ &  $-1$ & $-1$ & $+1$ & $X_0X_4$  \\   \hline  
$-1$ &  $-1$ & $-1$ & $-1$ & $X_3X_5$  \\   \hline  
\end{tabular}
\vspace{10pt}
\caption{ Lookup-table for the Z stabilizers of the surface-17 code.  }
\vspace{10pt}
\label{table:surface17Zstabs}
\end{table}

The lookup tables for the distance-3 BS code are shown in Tables \ref{table:BS17Xstabs} and \ref{table:BS17Zstabs} for the X and Z stabilizers, respectively.  Notice that these are equivalent to the lookup tables of the distance-3 repetition code.

\begin{table}
\centering
\vspace{10pt}
\begin{tabular}{ | c  c | c |}
\hline \; $S_X^{(1)}$ \; & \; $S_X^{(2)}$ \; & \; Error \;   \\ \hline \hline  
$+1$ &  $+1$ & $I$  \\   \hline  
$+1$ &  $-1$ & $Z_2$  \\   \hline  
$-1$ &  $+1$ & $Z_0$  \\   \hline  
$-1$ &  $-1$ & $Z_1$  \\   \hline  
\end{tabular}
\vspace{10pt}
\caption{ Lookup-table for the X stabilizers of the Bacon-Shor-17 code.  }
\vspace{10pt}
\label{table:BS17Xstabs}
\end{table}

\begin{table}
\centering
\vspace{10pt}
\begin{tabular}{ | c  c | c |}
\hline \; $S_Z^{(1)}$ \; & \; $S_Z^{(2)}$ \; & \; Error \;   \\ \hline \hline  
$+1$ &  $+1$ & $I$  \\   \hline  
$+1$ &  $-1$ & $X_6$  \\   \hline  
$-1$ &  $+1$ & $X_0$  \\   \hline  
$-1$ &  $-1$ & $X_3$  \\   \hline  
\end{tabular}
\vspace{10pt}
\caption{ Lookup-table for the Z stabilizers of the Bacon-Shor-17 code.  }
\vspace{10pt}
\label{table:BS17Zstabs}
\end{table}

\section{Readout infidelity of the ST qubit} \label{app:STqubit_readout_infidelity}

\begin{figure}
    \centering
    \includegraphics[width=0.95 \columnwidth]{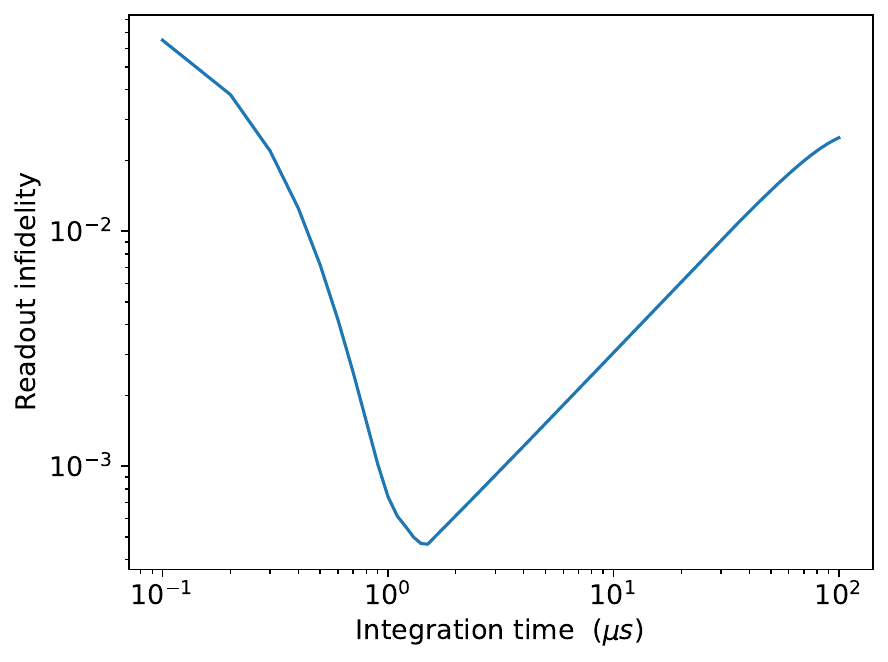} 
       \vspace{0pt}
    \caption{Readout infidelity of the ST qubit as a function of the integration time.}   
   \label{fig:ST_readout_infidelity}
\end{figure}

Figure \ref{fig:ST_readout_infidelity} shows the readout infidelity of the ST qubit as a function of the readout integration time.  We use this function in our simulations.  The integration time corresponding to the minimum readout infidelity in general differs from the integration time resulting in the lowest logical error rate.  This function was experimentally found in \cite{takeda2024rapid}.

\section{Impact of the duration of the CZ gate and the infidelity of the state preparation on the logical error rate} \label{app:othereffects}

\begin{figure*}[t]
    \centering
    \includegraphics[width=1 \textwidth]{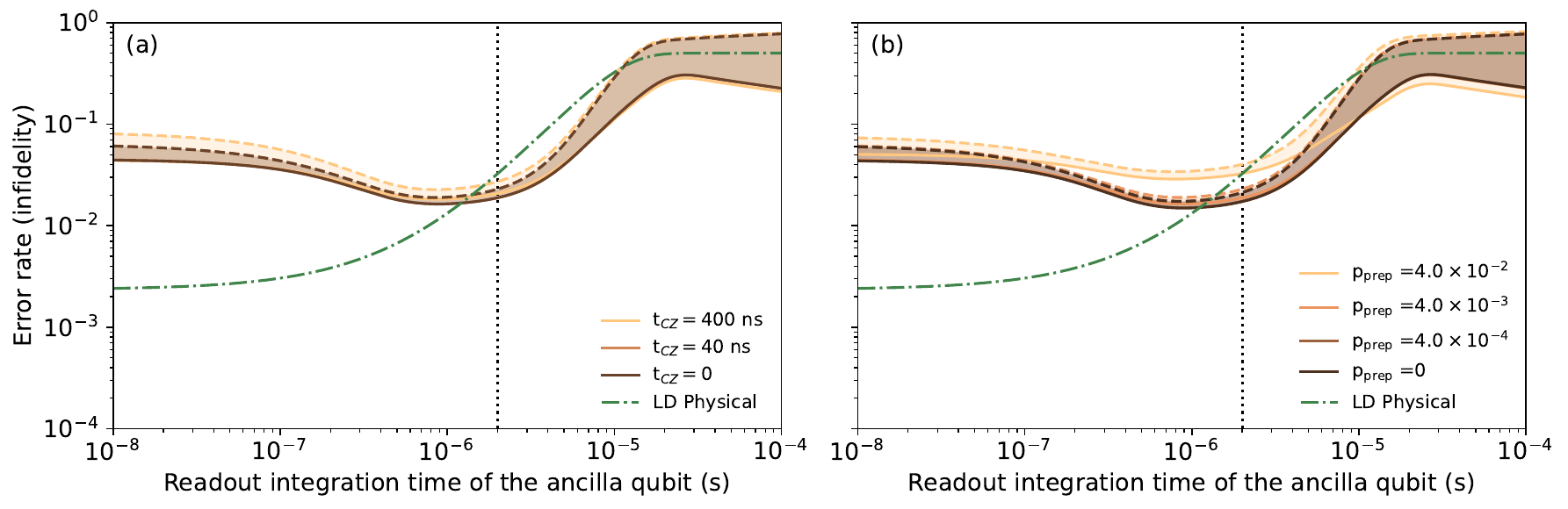} 
       \vspace{0pt}
    \caption{Logical error rate for 1 QEC step of the surface-17 in the hybrid encoding scheme as a function of the readout integration time of the ancillary qubits, for various (a) CZ gate durations and (d) state-preparation infidelities.  The ramping time is kept constant at $0.4~\mu $s.  In both panels, the dashed-dotted blue curves correspond to the infidelity of a physical qubit with $T_2^*=21~\mu$s that starts in the $|+ \rangle$ state and undergoes dephasing for the duration of the whole QEC step of the hybrid scheme.  The vertical dotted line marks the state-of-the-art integration time of $2.0 \ \mu$s.}   
    \label{fig:2params_appendix}
\end{figure*}

Figure \ref{fig:2params_appendix} shows the effect of the duration of the CZ gate and the state-preparation infidelity on the logical error rate of the surface-17 code in the hybrid encoding scheme.  The impact of these two parameters is rather negligible.

\textbf{Duration of the CZ gate:}
Since dephasing is the main source of noise in spin qubits, the duration of quantum operations is of critical importance.  However, as seen in Fig.~\ref{fig:2params_appendix}a, decreasing the duration of the CZ gate, while keeping the other parameters constant, will not have an impact on the logical error rate: the curve corresponding to the state-of-the-art value ($t_{\textrm{CZ}}=40 \, \textrm{ns}$) already overlaps with the expected curve for an instantaneous gate ($t_{\textrm{CZ}} = 0$).  This clearly indicates that the crucial step for memory errors is the long measurement duration.  Given the much longer duration of the measurement with respect to the CZ gate, optimizing the entangling gates' scheduling does not result in a noticeable improvement in the logical error rate, at least for the case analyzed here.

\textbf{Infidelity of the state preparation:} As seen in Fig.~\ref{fig:2params_appendix}b, decreasing the infidelity of the state preparation from the current value of $4.0 \times 10^{-3}$ by a factor $10$ has a negligible effect on the logical error rate.  This rather surprising behavior can be explained by the fact that, in our QEC protocol, state preparations only occur on the ancillary qubits.  The data qubits are initialized with a logical state already prepared.  Bit-flips after state preparations on the ancillary qubits do \textit{not} propagate to the data qubits; they only increase the effective readout error to: 

\begin{equation}
  p_{\textrm{readout,eff}} = p_{\textrm{readout}} + p_{\textrm{prep}} + O(p^2), \\
\end{equation}

where $p_{\textrm{readout}}$ is the nominal readout error rate, $p_{\textrm{prep}}$ is the state-preparation error rate, and $O(p^2)$ refers to quadratic powers of $p_{\textrm{readout}}$ and $p_{\textrm{prep}}$.  Since $p_{\textrm{readout}} \approx 4 \times 10^{-4}$ at the current integration time, when $p_{\textrm{prep}}$ is reduced from $4 \times 10^{-3}$ to $4 \times 10^{-4}$ the effective readout error rate goes from $p_{\textrm{readout,eff}} \approx 4.4 \times 10^{-3}$ to $p_{\textrm{readout,eff}} \approx 0.8 \times 10^{-3}$.  The fact that this drastic reduction in the effective readout error does not cause a significant reduction in the logical error rate indicates that the readout infidelity is not the dominant source of noise.  This behavior will most likely change under a QEC protocol where state preparations are present on the data qubits, as would occur in an actual experimental demonstration.  We leave this exploration for a future work.

\section{Effect of boundary-only initialization and readout on the logical error rates} \label{app:noreadoutbulk}

\begin{table}
\centering
\vspace{10pt}
\begin{tabular}{ | c | c |}
\hline \; Parameter \; & \; Value \;   \\ \hline \hline  
Phase-flip error rate per hop &  $4.0 \times 10^{-5}$ \cite{ShuttlingLieven} $^{\dagger}$   \\   \hline  
Bit-flip error rate per hop &  $1.08 \times 10^{-4}$ \cite{ShuttlingYoneda} $^{\dagger\dagger}$  \\   \hline  
Hop duration & $2$~ns \cite{ShuttlingLieven} $^{\ddagger}$ \\   \hline  
\end{tabular}
\vspace{10pt}
\caption{ Electron shuttling parameters employed in this section. $^{\dagger}$The phase-flip error rate is obtained from the reported conveyor-mode shuttling fidelity of $99.996\%$ \cite{ShuttlingLieven}.  $^{\dagger\dagger}$The bit-flip error rate is obtained from the reported polarization transfer fidelity of $99.9892\%$ \cite{ShuttlingYoneda}. $^{\ddagger}$The duration of an electron shuttling from one QD to a nearest-neighboring one is obtained from the reported shuttling speed of $10~\mu$m in $200$~ns and a QD-QD distance of $100$~nm \cite{ShuttlingLieven}.}
\vspace{10pt}
\label{table:ShuttlingParams}
\end{table}

In this section, we analyze the effect of having boundary-only initialization and readout, with the resulting need to have electron shuttling for the qubits in the bulk of the QEC code's layout.   Table \ref{table:ShuttlingParams} summarizes the parameters employed in the analysis.

\begin{figure*}[t]
    \centering
    \includegraphics[width=1 \textwidth]{surface_init.pdf} 
       \vspace{0pt}
    \caption{Electron shuttling sequence to initialize the data LD qubits and ancillary ST qubits on the surface-17 code.  The circles inside the dashed black perimeter correspond to the QDs that host the code's qubits (see Figure \ref{fig:layout_two_codes}b).  The larger green circles outside the perimeter correspond to charge sensors used to measure the ST qubits.  All LD data qubits (red) are initialized outside the perimeter and then shuttled to their corresponding QDs.}   
    \label{fig:shuttling_SurfCode}
\end{figure*}

\textbf{Effect on the logical state preparation of the surface-17 code:}  Figure \ref{fig:shuttling_SurfCode} shows the scheduling sequence to initialize the data LD qubits and ancillary ST qubits on the surface-17 code layout.  The LD data qubits (red) are initialized outside the perimeter and subsequently shuttled to their QDs.  The ST ancillary qubits do not need to be shuttled since they are all located along the boundary of the code's layout.  The number of necessary shuttling steps for the LD qubits ranges from 1 (for data qubits close to the boundary) to 3 (for the data qubit at the very center of the code).  The data qubits are initialized in the $Z$ basis, so only bit-flip errors are important.  Since the probability of a bit-flip error per hop is $1.08 \times 10^{-4}$ (see Table \ref{table:ShuttlingParams}), the electron shuttling increases the effective LD state-preparation error rate from $6.5 \times 10^{-3}$ to $(6.6 - 6.8) \times 10^{-3}$.  The rest of the operations remain the same because the readout of the ancillary ST qubits does not require any electron shuttling.  Moreover, since the shuttling speed is $2$ ns per hop, the extra waiting time is negligible.  Even after 3 shuttling steps, the extra waiting time is just $3 \times 2 $ ns $= 6$ ns, which is below the left-most limit of the $X$ axes in the main text figures ($10$ ns).     The state-preparation error rate increasing from $6.5 \times 10^{-3}$ to $6.8 \times 10^{-3}$ (worst case for the central data qubit) does not cause a qualitatively relevant change in the logical error rate curves.   

\textbf{Effect on the QEC step of the surface-17 code:}  Boundary-only initialization and readout has no effect on the QEC step of the surface code.  As shown in Figure \ref{fig:shuttling_SurfCode}, all the ancillary ST qubits are on the boundary of the code and have a charge sensor in close proximity.

\begin{figure*}[t]
    \centering
    \includegraphics[width=1 \textwidth]{bacon_shor_init.pdf} 
       \vspace{0pt}
    \caption{Electron shuttling sequence to initialize the data LD qubits on the BS-17 code.  As for the surface-17 code, the larger green circles outside the perimeter correspond to charge sensors used to measure the ST qubits.}   
    \label{fig:shuttling_PrepBSCode}
\end{figure*}

\textbf{Effect on the logical state preparation of the BS-17 code:}  As explained in the main text, the FT preparation of a $|0 \rangle _L$ or $|+ \rangle _L$ only involves the creation of a GHZ state along each row of the lattice.  For the distance-3 BS code, no verification is necessary, so the ancillary qubits are not used.  As shown in Figure \ref{fig:shuttling_PrepBSCode}, for the BS-17 code, we require not more than 1 electron hop.  Since the bit-flip error rate per hop is $1.08 \times 10^{-4}$, this implies that the effective LD qubit state-preparation error rate will increase from $6.5 \times 10^{-3}$ to $6.6 \times 10^{-3}$, at most.  As for the logical state preparation for the surface-17 code, this does not cause an appreciable change in the logical error rate curves from Figure \ref{fig:FTstateprep}.

\begin{figure*}[t]
    \centering
    \includegraphics[width=1 \textwidth]{bacon_shor_cycle4.pdf} 
       \vspace{0pt}
    \caption{Electron shuttling sequence to measure the ancillary ST qubits in the context of subsequent rounds of stabilizer measurements.  Only the two bulk ST ancillary qubits need to be shuttled to be read out by a charge sensor.  To minimize the errors, spin-to-charge conversion is performed before the shuttling. There are two possible charge configurations in the double QD: (0,2) and (1,1).}   
    \label{fig:shuttling_QECBSCode}
\end{figure*}


{\textbf{Effect on the QEC step of the BS-17 code:}  
In the BS-17 layout, two ST ancillary qubits are located in the bulk and therefore require shuttling for readout and initialization. The shuttling protocol is illustrated in Fig.~\ref{fig:shuttling_QECBSCode}. Crucially, we do not shuttle a coherent ST qubit. Instead, we first perform PSB locally in the bulk, converting the spin state into a charge configuration. After this spin-to-charge conversion, the double dot is in either a $(0,2)$ or $(1,1)$ configuration, where the first number refers to the outer dot (i.e., the QD closer to the array boundary). Only after this conversion do we shuttle the electron located in the outer dot toward a temporarily vacated site next to the charge sensor. If the charge configuration is $(0,2)$, the outer dot is empty and no electron is transferred. If it is $(1,1)$, one electron tunnels to the empty site. The boundary charge sensor then detects whether the occupation of that site has changed. In this way, the measurement outcome is obtained without transporting a coherent two-electron state.

Because the spin-to-charge conversion is completed before any transport occurs, there is no additional error contribution to the ancillary qubit beyond the standard PSB readout error. The only qubits affected by shuttling are the two edge data qubits that are temporarily displaced to open a transport path. Each of these data qubits undergoes two shuttling hops: one to vacate its original site and one to return. From Table~\ref{table:ShuttlingParams}, two shuttling steps lead to a phase-flip error probability of $2 \times 4 \times 10^{-5} = 8 \times 10^{-5}$ and a bit-flip error probability of $2 \times 1.08 \times 10^{-4} = 2.16 \times 10^{-4}$. For comparison, an LD qubit with $T_2^* = 21~\mu\text{s}$ that idles during the measurement of an ST qubit ($t = 2.4~\mu\text{s}$) acquires a phase-flip probability $p = \frac{1}{2}\left[1 - \exp\left(-(t/T_2^*)^2\right)\right] = 6.5 \times 10^{-3}$, which exceeds the shuttling-induced error by more than one order of magnitude. The additional waiting time associated with the shuttling operations is negligible ($2 \times 2~\text{ns} = 4~\text{ns}$) compared to the microsecond-scale measurement time.}

Taking into account the electron shuttling needed to account for boundary-only qubit initialization and readout results in a negligible increase in the logical error rate associated with the various protocols analyzed in the paper.  We notice that this will only apply to the distance-3 QEC codes studied here.  For higher-distance codes, electron shuttling will probably have appreciable effects on the overall performance of QEC protocols.

\bibliography{PRA_Resubmission/leak_STZ_resub}

\end{document}